\documentclass[fleqn,usenatbib]{mnras}

\usepackage{newtxtext,newtxmath}
\usepackage[T1]{fontenc}
\usepackage{ae,aecompl}

\usepackage{graphicx}	
\usepackage{dblfloatfix}
\usepackage{subcaption}
\usepackage{amsmath}	
\usepackage{amssymb}	
\usepackage{color}
\usepackage[dvipsnames]{xcolor}
\usepackage[normalem]{ulem}
\usepackage{booktabs}
\usepackage{caption}

\newcommand{\hst}{\textit{HST}}
\newcommand{\hstlong}{\textit{Hubble Space Telescope}}

\newcommand{\xmmlong}{\textit{XMM-Newton}}
\newcommand{\xmm}{\textit{XMM}}
\newcommand{\rosat}{{\em ROSAT}}
\newcommand{\nustar}{\textit{NuSTAR}}
\newcommand{\nicer}{{\em NICER}}

\newcommand{\nh}{\mbox{$N_{\rm H}$}}

\newcommand{\teff}{\mbox{$T_{\rm eff}$}}
\newcommand{\tinfeff}{\mbox{$T_{\rm eff}^{\infty}$}}
\newcommand{\rinfty}{\mbox{$R_{\infty}$}}
\newcommand{\rns}{\mbox{$R_{\rm NS}$}}
\newcommand{\mns}{\mbox{$M_{\rm NS}$}}

\newcommand{\chisq}{\mbox{$\chi^2$}}

\newcommand{\Msun}{\mbox{$M_\odot$}}

\newcommand{\xray}{\mbox{X--ray}}

\newcommand{\simlt}{\mathrel{\hbox{\rlap{\hbox{\lower4pt\hbox{$\sim$}}}\hbox{$<$}}}}
\newcommand{\simgt}{\mathrel{\hbox{\rlap{\hbox{\lower4pt\hbox{$\sim$}}}\hbox{$>$}}}}

\newcommand{\ee}[1]{\mbox{$10^{#1}$}}
\newcommand{\tee}[1]{\mbox{$\times 10^{#1}$}}

\newcommand{\unit}[1]{\mbox{$\rm\,#1$}}

\newcommand{\G}{\mbox{$\,G$}}
\newcommand{\msun}{\mbox{$\,M_\odot$}}

\newcommand{\km}{\hbox{$\,{\rm km}$}}
\newcommand{\K}{\hbox{$\,{\rm K}$}}

\newcommand{\keV}{\mbox{$\,{\rm keV}$}}
\newcommand{\eV}{\mbox{$\,{\rm eV}$}}

\newcommand{\msec}{\mbox{$\,{\rm ms}$}}
\newcommand{\yr}{\mbox{$\,{\rm yr}$}}

\newcommand{\pc}{\mbox{$\,{\rm pc}$}}
\newcommand{\persec}{\mbox{$\,{\rm s^{-1}}$}}

\newcommand{\percmsq}{\mbox{$\,{\rm cm^{-2}}$}}

\newcommand{\cgsaccel}{\mbox{$\,{\rm cm\,s^{-2}}$}}

\title[NS radius measurement for  PSR~J0437--4715]{Neutron star radius measurement  from  the ultraviolet and soft X--ray thermal emission of  PSR~J0437--4715}

\author[Gonz\'{a}lez-Caniulef et al.]{
Denis Gonz\'alez-Caniulef$^{1, 2}$\thanks{E-mail: denis.caniulef.14@ucl.ac.uk},
Sebastien Guillot$^{3, 4}$,
and Andreas Reisenegger$^{2}$
\\
$^{1}$ Mullard Space Science Laboratory, University College London, Holmbury St. Mary, Dorking, Surrey, RH5 6NT, UK\\
$^{2}$ Instituto de Astrof\'{i}sica, Facultad de F\'{i}sica,  Pontificia 
   Universidad Cat\'{o}lica de Chile, Av. Vicu\~{n}a Mackenna 4860,\\ 8970436
   Macul, Santiago, Chile\\
$^{3}$ IRAP, CNRS, 9 avenue du Colonel Roche, BP 44346, F-31028 Toulouse Cedex 4, France\\
$^{4}$ Universit\'{e} de Toulouse, CNES, UPS-OMP, F-31028 Toulouse, France
}

\date{Accepted XXX. Received YYY; in original form ZZZ}

\pubyear{2015}

\begin{document}
\label{firstpage}
\pagerange{\pageref{firstpage}--\pageref{lastpage}}
\maketitle

\begin{abstract}
We analyzed the thermal emission from the entire surface of the millisecond pulsar PSR~J0437--4715 observed in the ultraviolet and soft X--ray bands. For this, we calculated non-magnetized, partially ionized atmosphere models of hydrogen, helium, and iron compositions and included plasma frequency effects that may affect the emergent spectrum. This is particularly true for the coldest atmospheres composed of iron (up to a few percent changes in the soft X--ray flux).  Employing a Markov chain Monte Carlo method, we found that the spectral fits favour a hydrogen atmosphere, disfavour a helium composition and rule out iron atmosphere and blackbody models. By using a Gaussian prior on the dust extinction, based on the latest 3D map of Galactic dust, and accounting for the presence of hot polar caps found in previous work, we found that the hydrogen atmosphere model results in a well-constrained neutron star radius $\rns = 13.6^{+0.9}_{-0.8}\km$ and bulk surface temperature $\tinfeff=\left(2.3\pm0.1\right)\tee{5}\K$. This relatively large radius favours a stiff equation of state and disfavours a strange quark composition inside neutron stars.
\end{abstract}


\section{Introduction} 

The study of the thermal emission from neutron stars (NSs) is particularly relevant to understand their cooling history, determine their radii, and constrain the equation of state (EOS) of ultra-dense matter. In general, the thermal emission from NSs is expected to be reprocessed by an atmosphere, whose spectrum depends on the temperature gradient, surface composition, and magnetic field strength of the source. There already exists substantial literature dedicated to the effects of these parameters on the emergent spectra of relatively hot NS atmospheres, with temperature $T>10^{6}$~K \citep[for reviews, see e.g.,][]{zavlin07,ozel13,potekhin14}.

The spectra of cooler NS atmospheres, in which plasma effects start to become important, have been less studied. However, these atmosphere models are relevant for analyzing the thermal emission from old NSs (ages $>\ee{6}\yr$), such as millisecond pulsars (MSPs), which have lost most of their thermal energy.  Up to now, \hstlong\ (\hst) observations have revealed ultraviolet (UV) emission from two MSPs, PSR~J0437--4715 \citep[hereafter ``J0437'';][]{kargaltsev04,durant12} and PSR~J2124--3358 \citep[``J2124'';][]{rangelov17}, and from the middle-aged classical pulsar PSR~B0950+08 \citep[``B0950'';][]{pavlov17}. In all three cases, the interpretation of the thermal spectrum as blackbody (BB) emission yields bulk surface temperatures around $\ee{5}\K$, whereas an upper limit of $4\times\ee{4}\K$ was inferred from the non-detection of the old, very slow pulsar PSR~J2144--3933 \citep{guillot19}. More realistic estimates of the surface temperature of these objects, using atmosphere models, will help to distinguish between different possible heating mechanisms in old NSs \citep{gonzalez10}. This would in turn provide constraints on NS internal parameters, such as the superfluid energy gaps, which regulate the strength of these mechanisms \citep{petrovich10,petrovich11,gonzalezjimenez15}. 

Furthermore, J0437 and J2124, as well as PSR~J0030+0451 (``J0030''), are among the targets of the \emph{Neutron Star Interior Composition Explorer} mission (\nicer, \citealt{gendreau16,gendreau17}). \nicer\ aims at measuring the mass \mns{} and radius \rns{} of these MSPs through the effect of gravitational light bending and other relativistic effects on their X--ray light curves produced by hot polar caps \citep{bogdanov13,ozel16,miller16}.  In fact, combined analysis of hydrogen (H) atmosphere models with \xmmlong{} X--ray spectral/timing observations of the hot polar cap emission have permitted establishing some constraints on the radius of J2124, J0030, and J0437, which, assuming a NS mass $\mns=1.4\msun$, resulted in $\rns>7.8\km$ (68\% confidence, \citealt{bogdanov08}), $\rns>10.4\km$ (99\% confidence, \citealt{bogdanov09}) and $\rns>10.9\km$ (3$\sigma$, confidence, \citealt{bogdanov13}), respectively.  In light of upcoming \nicer{} analyses, it is therefore important to characterize the atmospheric properties of MSPs (e.g., composition and temperature) as accurately as possible to establish further constraints on the properties of these sources.

We model non-magnetized, partially ionized NS atmospheres for temperatures down to $\sim 10^{4.5}$~K, and fit them to the UV and soft X--ray spectra of J0437. Besides being the main target for \nicer, J0437 is the brightest and nearest MSP, with a precisely measured distance $d=156.79\pm0.25\pc$ \citep{reardon16}. In addition, it is in a 5.74 day binary orbit with a helium-core white dwarf companion \citep{bailyn93}, allowing for a precise radio-timing measurement of the pulsar mass, $\mns=1.44\pm0.07\msun$ \citep{reardon16}.  The white dwarf has an effective temperature of $3950\pm150\K$ \citep{durant12}, making it very unlikely to contribute significantly to the UV emission of the system. The spin period of J0437, $P=5.76\msec$, and its spin-down rate, $\dot{P}=5.73\tee{-20}\unit{s\persec}$, imply a large spin-down age (after kinematic corrections), $\tau=P/2\dot{P}=6.7\unit{Gyr}$, and a weak dipole magnetic field, $B=2.8\tee{8}\G$.

The X--ray spectrum of J0437 is composed of two thermal components from the pulsar's hot polar caps \citep{zavlin98}, generally fitted with a pair of NS atmosphere components (with $T\sim\ee{6}\K$), and a non-thermal component fitted with a power-law \citep{zavlin02b}. Because of its proximity, J0437 is also the only MSP for which the $\sim\ee{5}\K$ thermal emission from the entire pulsar surface is detectable in the soft X--ray range below $\sim0.4\keV$. This third (cool) thermal component is clearly seen in the UV \citep{kargaltsev04,durant12}, but it was poorly constrained in studies that only considered the X-ray data, due to uncertain contributions of a non-thermal component \citep{bogdanov13}. This is because the mild excess below $\sim0.4\keV$ in the \xmm\ data could be compensated by a soft power-law. More recent work used \nustar\ observations to better constrain the high-energy tail (at $\gtrsim 4\keV$), which lifted ambiguities with the spectral modeling at lower energies ($\lesssim 0.4\keV$). Combined \rosat, \xmmlong\ and \nustar\ spectral analysis confirmed the presence of this third thermal component (with $T\sim\ee{5}\K$, fitted with a BB), which was interpreted as the thermal emission coming from the entire surface \citep{guillot16a}.

However, only the Wien tail of this cool surface emission is detected in the soft X--ray regime. Therefore, its minimal contribution to the X--ray flux in the 0.1--0.5 keV range, affected by absorption due to the interstellar medium, and dominated by the hotter thermal components, prevented precisely determining the surface temperature and emitting area (i.e., the NS radius).  In addition to the evidence for the cool thermal emission in the soft X--rays, J0437 is also the only pulsar for which one can precisely determine, thanks to the pulsar's proximity, the slope of this emission in the UV band with spectroscopic observations \citep{kargaltsev04,durant12}. 

This paper aims at obtaining better constraints on the cool thermal emission of J0437 by combining the UV and soft X--ray observations. We model and apply realistic NS atmosphere models, for various compositions, to the UV data from \hst\ and soft X--ray data from \rosat. The organization of the article is as follows. In Section~\ref{sec:theo}, we describe the theoretical framework to compute the emergent spectrum from a NS atmosphere, and how we introduce the plasma effects. In Section~\ref{sec:model}, we verify the accuracy of our models and investigate their properties, particularly the plasma effects. In Section~\ref{sec:applications}, we confront our atmosphere models for different compositions to the observed UV to soft X--ray spectral energy distribution of J0437. A summary of our main conclusions and the discussion are given in Section~\ref{sec:ccl}.

\section{Theoretical framework}
\label{sec:theo}

Substantial literature on both magnetic and non-magnetic NS atmosphere models already exists \citep[for reviews see][]{zavlin07,ozel13,potekhin14}. In particular, spectra from non-magnetized, passively cooling NSs are usually obtained via the computation of the atmosphere structure coupled with either: a) the Milne integral \citep{romani87,rajagopal96,pons02}, b) the radiative transfer equation in the form of a second-order boundary problem \citep{zavlin96,heinke06b,suleimanov07,haakonsen12}, or c) the radiative transfer equation  using the Rybicki method \citep{gansicke02}.

\subsection{Neglect of magnetic effects}
\label{sec:magnetic}

The criteria to establish whether non-magnetized atmospheres are suitable for analyses of the thermal emission from different classes of NSs consider the temperature, spectral energy range, and magnetic field strength of the source. Basically, a magnetic field changes the properties of the atmosphere in two ways: by modifying the energy levels of the atoms, which changes the bound-bound and bound-free opacities, and by modifying the dynamics of free electrons with kinetic energy below the electron cyclotron energy, which changes the free-free opacities.  Considering the electron cyclotron energy $E_\mathrm{c}= \hbar eB/m_{e}c\approx 1\, B_8\unit{eV}$, magnetic fields are negligible for the bound-bound and bound-free opacities if the ratio $E_\mathrm{c}/Z^2\mathrm{Ry}\sim 0.1\, B_8 \,Z^{-2}\ll1$, where $Z$ is the atomic number,  $\mathrm{Ry}=13.6\unit{eV}$ is the Rydberg energy, and  $B_8=B/10^8\G$. Similarly, for the free-free  opacities, magnetic fields are negligible if the ratio $E_\mathrm{c}/k_{B}T\sim 0.1\, B_8/T_5 \ll 1$, with $T_5=T/10^5\K$, or the spectral energy range of interest is above the electron cyclotron energy, $E\gg E_\mathrm{c}\approx 1\, B_8\unit{eV}$.   

Spectra for \emph{fully ionized} atmosphere models of NSs with different field strengths have been reported, for example, in Figure 2.7 of \cite{lloyd03}. In particular, for the lowest temperature considered in that work, $\log(T/\mathrm{K}) = 5.6$, the spectrum for  $B=\ee{8}\unit{G}$ is indistinguishable from that of a non-magnetic NS. As also shown in the same figure, the magnetic field produces an absorption feature around the electron cyclotron energy. In the case of J0437,  the magnetic field is  $B_8\approx 2.8$, the associated electron cyclotron frequency is $\log(E_\mathrm{c}/\mathrm{keV}) = -2.6$, and we will fit the spectrum to UV data well above this value, for $\log(E/\mathrm{keV}) > -2.2$, finding typical temperatures $T_5\sim 3$.

Depending on the composition, \emph{partially ionized} atmospheres show absorption features at different energies. In particular, the H atmosphere spectrum has a Lyman alpha absorption feature at $E=10.2\eV$, which, depending on the NS gravitational redshift, can be within the range of the ultraviolet \hst{} observations. Magnetic fields such as those present in MSPs are strong enough to induce a large Zeeman effect, i.e., the splitting of the Lyman alpha absorption feature into three separate components \citep[see e.g.,][]{kargaltsev04}. These are expected to be washed out in the measured, phase-averaged spectrum, because the latter combines radiation from different parts of the NS surface, where the magnetic field strength and direction are expected to be very different, thus placing the absorption components at different wavelengths. 
Therefore, we will ignore the presence of the magnetic field in our atmosphere model calculations. For the spectral fitting of J0437 (Section \ref{sec:fits}), we eliminate the Lyman alpha absorption feature from the spectra of the H atmosphere models by linearly interpolating through the spectral range covered by this feature.

\begin{figure*}
    \includegraphics[width=0.342\textwidth, viewport=15 5 525 390,clip]{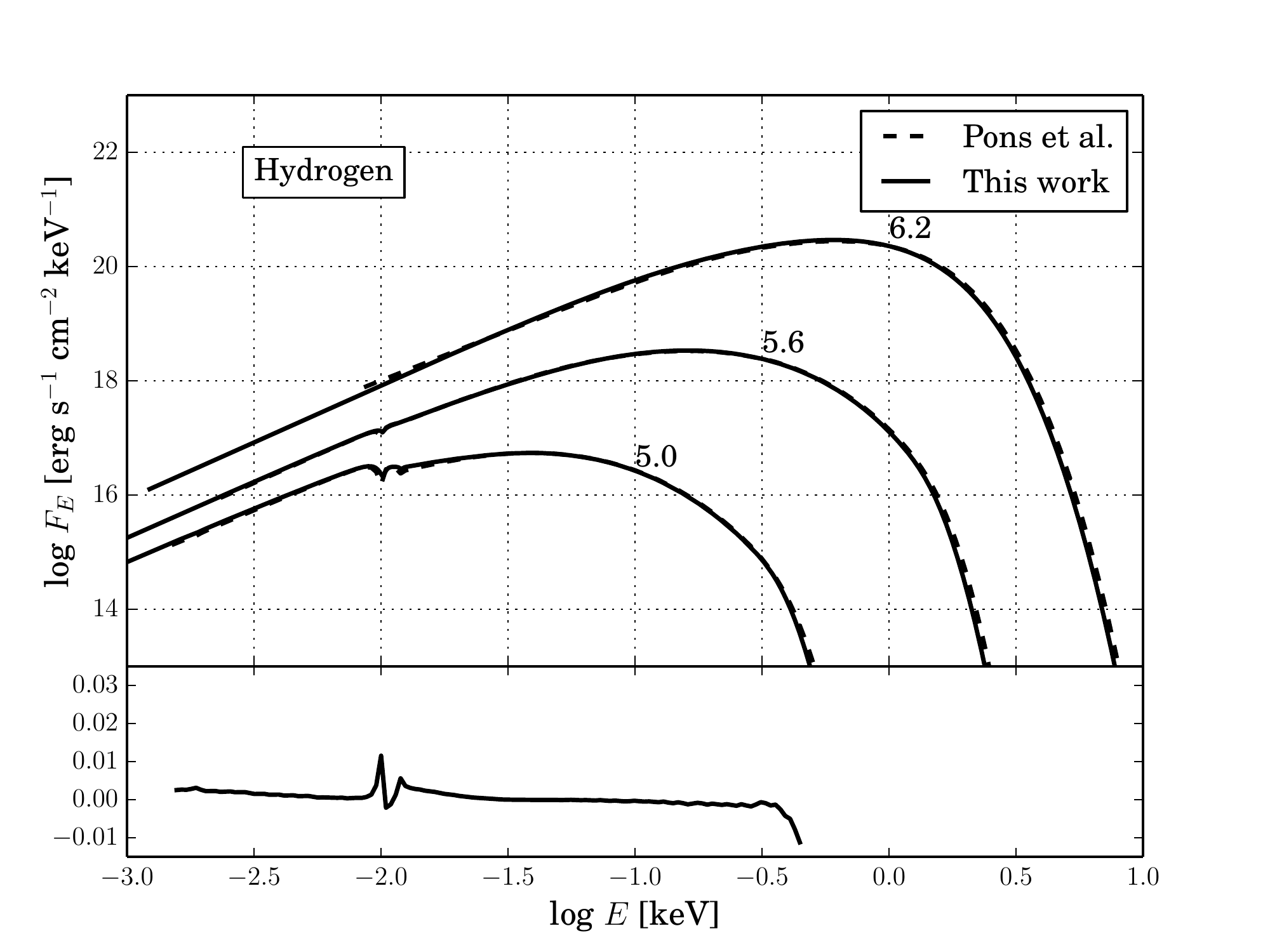}
    \includegraphics[width=0.322\textwidth, viewport=45 5 525 390,clip]{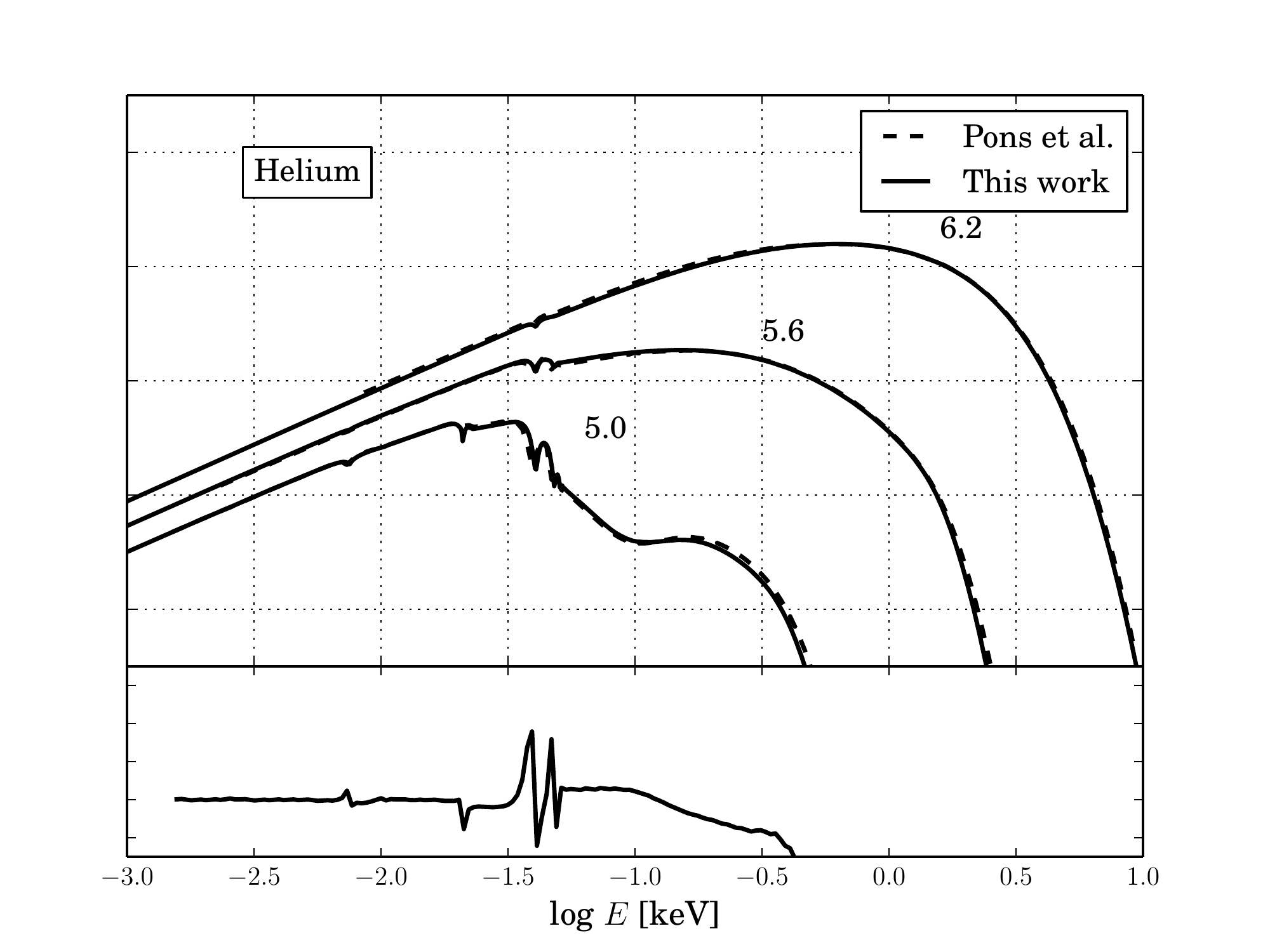}
    \includegraphics[width=0.322\textwidth, viewport=45 5 525 390,clip]{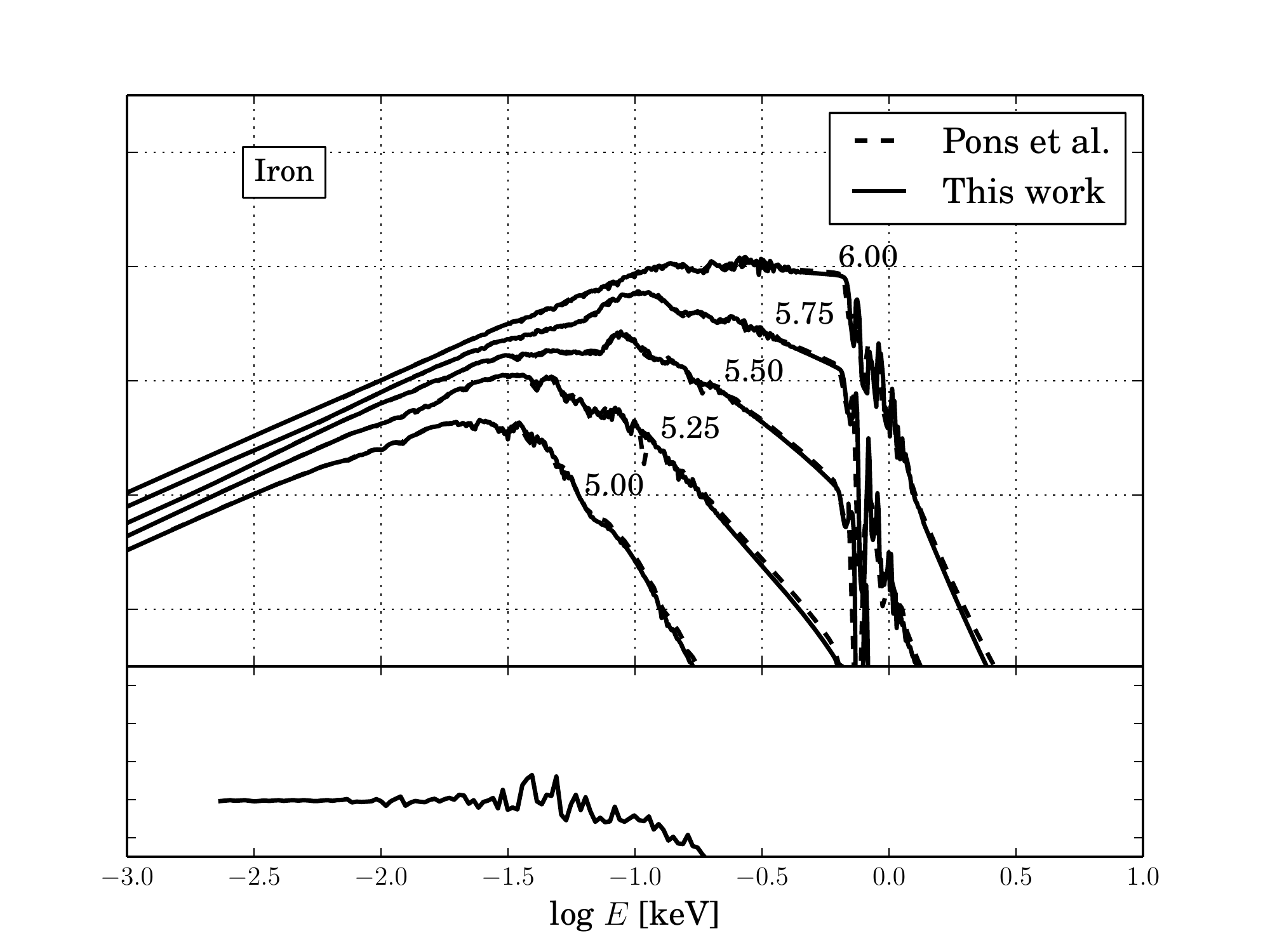}
    \caption{Spectra for non-magnetic NS atmospheres with pure H, He, and Fe composition. All spectra are computed for a NS surface gravity $g_{\mathrm{eff}}=2.43\tee{14}\cgsaccel$. The logarithms of the effective NS surface temperatures are labeled in each spectrum. In all panels, the dashed lines correspond to the spectra computed by \citet{pons02}, and the solid lines correspond to the spectra computed in the present work. The lower panels show the fractional difference, $\left(f_{0}-f_{1}\right)/f_{0}$, between the fluxes of this work, $f_{0}$, 	and those of \citet{pons02}, $f_{1}$, for NS atmospheres with effective surface temperature $\teff=\ee{5.0}\K$.}
	\label{fig:test}
\end{figure*}

Another important effect of magnetic fields on NS atmospheres is that they can suppress convective instabilities. As shown by \cite{rajagopal96}, a pure iron (Fe) atmosphere with $T\sim\ee{5}\K$ is unstable to convective motion in zones of the atmospheres with optical depths $\tau\sim0.1-1.0$, which could modify the temperature gradient of the atmosphere and produce a dramatic effect in the emergent spectra. However, they also showed that a magnetic field $B \gtrsim \ee{7}\G$ can suppress this instability and, therefore, it should not be present in the atmospheres of MSPs. Consistently, our models do not include convection.

\subsection{Atmosphere model calculations}
\label{sec:atmcalc}

We use our own new code based on the iterative scheme discussed in  \citet[][see also \citealt{rajagopal96,pons02}]{romani87} to simultaneously calculate the atmosphere structure and the spectral energy distribution via the Milne integral, for the case of unmagnetized NSs with low temperatures. It imposes that the NS atmosphere is in hydrostatic and radiative equilibrium. The latter means that the radiative flux through the atmosphere is constant and there is no additional source of energy. Because the thickness of the atmosphere is much smaller than the radius of the star, the radiative transfer equation is solved in the plane-parallel approximation, assuming the atmosphere is in local thermodynamic equilibrium. 

In order to determine the structure of the atmosphere, we solve the equation of hydrostatic equilibrium 
\begin{equation}
\label{eq1}
\frac{\mathrm{d}P}{\mathrm{d}\tau_\mathrm{R}} = \frac{g_{\mathrm{eff}}}{\kappa_\mathrm{R}}, 
\end{equation}
where $P$ is the pressure, $\kappa_\mathrm{R}$ is the Rosseland mean opacity (defined later in equation \ref{eq:rosseland}) and $\mathrm{d}\tau_\mathrm{R}=\rho\kappa_\mathrm{R}\,\mathrm{d}\ell$, with $\rho$ and $\ell$ the density and physical depth of the atmosphere, respectively. Here, for a given mass \mns\ and coordinate radius \rns, the gravitational acceleration $g_{\mathrm{eff}}= \left(1+z\right)G\mns/\rns^2$ and the gravitational redshift $1+z=(1-2G\mns/\rns\, c^2 )^{-1/2}$, where $G$ and $c$ correspond to the gravitational constant and the speed of light, respectively. For equation (\ref{eq1}), we use the boundary condition $P(\tau_\mathrm{R}=0)=0$ and an ideal gas equation of state, adding the pressure of degenerate electrons, which becomes relevant in the deepest zones of the atmosphere.

The problem is solved through successive iterations, starting from an initial temperature profile given by the solution for the gray atmosphere, $T^{4} = (3/4)\, \teff^4 (\tau_\mathrm{R} + q)$, where \teff{} is the effective temperature and $q = 2/3$. Subsequently, we calculate the energy-dependent flux through the atmosphere. Since the absorptive opacity $\kappa_{\scriptscriptstyle E}^\mathrm{a}\sim\ee{3}-\ee{5}\unit{cm^{2}\,g^{-1}}$ is much larger than the electron scattering opacity $\kappa^\mathrm{sc}\approx 0.1-0.2\unit{cm^{2}\,g^{-1}}$, we neglect the electron scattering effects. In this way, the expression for the energy-dependent flux reduces to the Milne integral \citep{mihalas78},
\begin{align}
	\begin{split}
	\label{eq7}
	F_{E}\left(\tau_{\scriptscriptstyle E}\right)=&2\pi\left[\int_{\tau_{\scriptscriptstyle E}}^{\infty}S_{\scriptscriptstyle E}\left(\tau_{\scriptscriptstyle E}^\prime \right)E_2\left(\tau_{\scriptscriptstyle E}^\prime-\tau_{\scriptscriptstyle E}\right)\mathrm{d}\tau_{\scriptscriptstyle E}^\prime\right.\\
				&-\left.\int_0^{\tau_{\scriptscriptstyle E}}S_{\scriptscriptstyle E}\left(\tau_{\scriptscriptstyle E}^\prime\right)E_2\left(\tau_{\scriptscriptstyle E}-\tau_{\scriptscriptstyle E}^\prime\right)\mathrm{d}\tau_{\scriptscriptstyle E}^\prime\right],
	\end{split}
\end{align}
where the source function $S_E(x)$ is just the Planck function, 
\begin{equation}
E_2(x) = \int_1^\infty \frac{e^{-xt}}{t^2}\mathrm{d}t
\end{equation}
is the second exponential integral, and 
\begin{equation}
\tau_{\scriptscriptstyle E} = \int_0^{\tau_R}\frac{\kappa_{\scriptscriptstyle E}}{\kappa_R}\mathrm{d}\tau_R^\prime
\end{equation}
gives the transformation from Rosseland mean to energy-dependent optical depths. 

Finally, in order to obtain a specified, constant energy-integrated radiative flux through the atmosphere, $F=\sigma\teff^{4}$, we apply the Lucy-Uns\"{o}ld correction to the temperature profile, which is given by 
\begin{align}
\begin{split}
\Delta T\left(\tau\right)=&\frac{1}{16\sigma T\left(\tau\right)^3} \left[ \frac{\kappa_J}{\kappa_P}\left( 3\int_0^\tau \frac{\kappa_F\left(\tau^\prime\right)}{\kappa_R\left(\tau^\prime\right)}\Delta F\left(\tau^\prime\right) \mathrm{d}\tau^\prime + 2\Delta F\left(0\right)\right)\right.\\ 
&- \left.\frac{\kappa_R}{\kappa_P}\frac{\mathrm{d}\Delta F\left(\tau\right)}{\mathrm{d}\tau_\mathrm{R}} \right], 
\end{split}
\end{align}
where $\Delta F$ is the departure from the specified, constant flux $F$. In the previous expression, the quantities  
\begin{equation}
\kappa_J \equiv \int_0^\infty \kappa_{\scriptscriptstyle E}^\mathrm{a} J_{\scriptscriptstyle E} \mathrm{d}E\big/J,
\end{equation}
\begin{equation}
\kappa_P \equiv \int_0^\infty \kappa_{\scriptscriptstyle E}^\mathrm{a} B_{\scriptscriptstyle E} \mathrm{d}E\big/B,
\end{equation}
\begin{equation}
\kappa_F \equiv \int_0^\infty (\kappa_{\scriptscriptstyle E}^\mathrm{a} +\kappa^\mathrm{sc}) F_{\scriptscriptstyle E} \mathrm{d}E\big/F,
\end{equation}
and 
\begin{equation}
\frac{1}{\kappa_\mathrm{R}} \equiv \int_0^\infty \frac{1}{\kappa_{\scriptscriptstyle E}^\mathrm{a} +\kappa^\mathrm{sc}} \frac{\mathrm{d}B_{\scriptscriptstyle E}}{\mathrm{d}T}\mathrm{d}E\bigg/ \frac{\mathrm{d}B}{\mathrm{d}T}
\label{eq:rosseland}
\end{equation}
are the absorption mean, Planck mean, flux mean, and Rosseland mean opacities, respectively \citep{mihalas78}. Here, $J$ and $B$  are the mean intensity and Planck function integrated in energy, respectively, and we approximate
$\kappa_J=\kappa_P$. A relatively constant flux (error $\lesssim 1\%$) is reached in $\sim 15$ iterations.  In this procedure, we take into account the corrections from General Relativity in the emergent spectrum. This means that the flux measured by an observer at distance $D$ is 
\begin{equation}
F_E^\infty\left(E\right) = \frac{F_{E}\left(\left[1+z\right]E\right)}{1+z} \left(\frac{\rns}{D}\right)^2,
\end{equation}
where $F_{E}\left(\left[1+z\right]E\right)$ is the flux at the NS surface.

In order to compute the emergent spectrum, we use 100 energy bins logarithmically spaced from $\ee{-4}\keV$ to $10\keV$ and a grid of 120 depth levels logarithmically spaced in Rosseland optical depth, $\tau_R$, from $\ee{-3}$ to $\ee{3}$. Once the proper atmosphere structure is iteratively obtained, the spectrum is calculated using a denser grid with 900 energy bins. We use the energy-dependent opacities and the Rosseland and Planck mean opacities for H, helium (He), and Fe from the Los Alamos Opacity Project\footnote{\url{http://aphysics2.lanl.gov/cgi-bin/opacrun/tops\_txt.pl}} (LANL; \citealt{magee95}), which include bound-bound, bound-free, and free-free transitions.
The LANL opacity tables also provide the number of free electrons per nucleus for a given composition, temperature and density
\citep[for details about ionization calculations see][and references therein]{magee95}.
However, the tables do not cover completely the energy-dependent opacities for relatively low energies, $E\sim \ee{-4}-\ee{-2}\keV$. We complete this region using the free-free opacity, which is dominant\footnote{This may not be strictly true for a heavy element composition, such as Fe atmospheres. However, we use the free-free opacity in an energy range with a relatively low radiative flux, where the atmosphere is optically thick, and the photosphere has a relatively small temperature gradient.  This means that the emergent spectrum is largely unchanged  by increasing the opacities, for example, due to additional bound-bound or bound-free transitions.}
in this range and is given, in CGS units, by \cite{rybicki79} as:
\begin{equation}
\kappa_{\nu}^{\rm ff} = 3.7 \tee{8}\,T^{-1/2}  n_{e}^2 \sum_{i} n_{i} Z_{i}^2 \nu^{-3}\left(1-e^{-h\nu/kT}\right) \bar g_{\nu}^{\rm ff}(\nu,T),
\end{equation}
where $i$ labels the kind of ions, $Z_i$ is the charge of the ions, $n_i$ is the ion number density, and $\bar g_{\nu}^{\rm ff}$ is the Gaunt factor, also obtained from \cite{rybicki79}.

\subsection{Plasma effects} 
\label{sec:plasma}

In the UV range, the emergent spectrum can be affected by absorption features due to atomic transitions in this energy range, as well as by plasma effects. The latter can be seen  through the standard expression for the plasma frequency
\begin{equation}
\label{eq:plasma_freq}
\omega_p = \left(\frac{4\pi e^2 n_e}{m_e}\right)^{1/2},
\end{equation}
where $-e$ is the electron charge, $n_e$ is the electron number density, and $m_e$ is the electron mass. This frequency can be estimated by combining the equation of hydrostatic equilibrium (Equation~\ref{eq1}) with an ideal gas equation of state $P = n k_{B} T$, where $n$ is the particle density and $k_B$ is the Boltzmann constant, yielding $n \sim g_\mathrm{eff}\tau/\left(\kappa_{R} k_{B}T\right)$. Assuming $n_{e} \sim n$, an optical depth $\tau\sim 1$, a H Rosseland mean opacity $\kappa_{R}\sim\ee{4}\unit{cm^{2}\,g^{-1}}$, and an effective surface gravity $g_{\mathrm{eff}} \sim 10^{14}\cgsaccel$, we obtain $\hbar \omega_{p} \sim\hbar\left[4\pi e^2 g_\mathrm{eff}\tau/\left(m_{e}\kappa_{R} k_{B}T\right)\right]^{1/2}\sim1\eV$, which is close to the UV range and therefore may affect the analysis of HST observations of cool NSs. 

In a plasma, the dispersion relation connecting the wave number, $k$, the frequency, $\omega$, and the plasma frequency, $\omega_p$, is given by:
\begin{equation}
    \omega = \left(\omega_{p}^2 + c^2 k^2\right)^{1/2},
\end{equation}
where $c$ is the speed of light. Since for $\omega < \omega_p$ the wave number becomes imaginary, $\omega_p$ defines a cutoff frequency below which there is no electromagnetic wave propagation in the plasma. \cite{aharony79} showed that, for $\omega > \omega_p$, the frequency-dependent opacities $\kappa_{\omega}$ should be replaced by:
\begin{equation}
    \kappa_{\omega} \rightarrow \frac{\kappa_{\omega}}{1-(\omega_p/\omega)^{2}}.
\end{equation}
In other words, the frequency-dependent opacities tend to infinity when a photon with frequency just above $\omega_p$ passes through the plasma. Thus, this correction to the frequency-dependent opacity has the effect, in the radiative transfer equation, of blocking the flux of photons with $\omega < \omega_{p}$ through the plasma. We therefore incorporate these considerations into our NS atmosphere model calculation and quantify their effects, before applying these models to the UV/X--ray data of J0437.

\section{Model comparison and results}
\label{sec:model}

In order to test our atmosphere calculation code, we generate spectra for different temperatures and compositions and compare them with the spectra of \cite{pons02} for H and He atmospheres with temperatures ranging from  $\teff=\ee{5.0}$ to $\ee{6.2}\K$, and for Fe atmospheres with effective temperatures ranging from $10^{5.0}$ to $10^{6.0}$~K. Like the present work, \cite{pons02} follow a standard  technique to model the NS atmosphere \citep{romani87,rajagopal96} and compute the emergent spectrum using the LANL opacities.  The main differences are that their calculations consider only 200 energy bins, compared to 900 in our case, and cover the range of Rosseland optical depths $\ee{-8}<\tau_R<\ee{2}$, whereas we used $\ee{-3}<\tau_R<\ee{2}$ for the comparison and $\ee{-3}<\tau_R<\ee{3}$ for all other calculations. The range $\ee{-8}<\tau_R<\ee{-3}$, which we do not cover, does not make a significant difference because very few photons are emitted or absorbed in this region. Including the interval $\ee{2}<\tau_R<\ee{3}$, on the other hand, slightly increases the flux in the high-energy tail (worsening the agreement with \citealt{pons02}, as expected), but does not noticeably affect most of the spectrum.

\begin{figure}
\centering
\includegraphics[width=\hsize, viewport=15 10 520 395,clip]{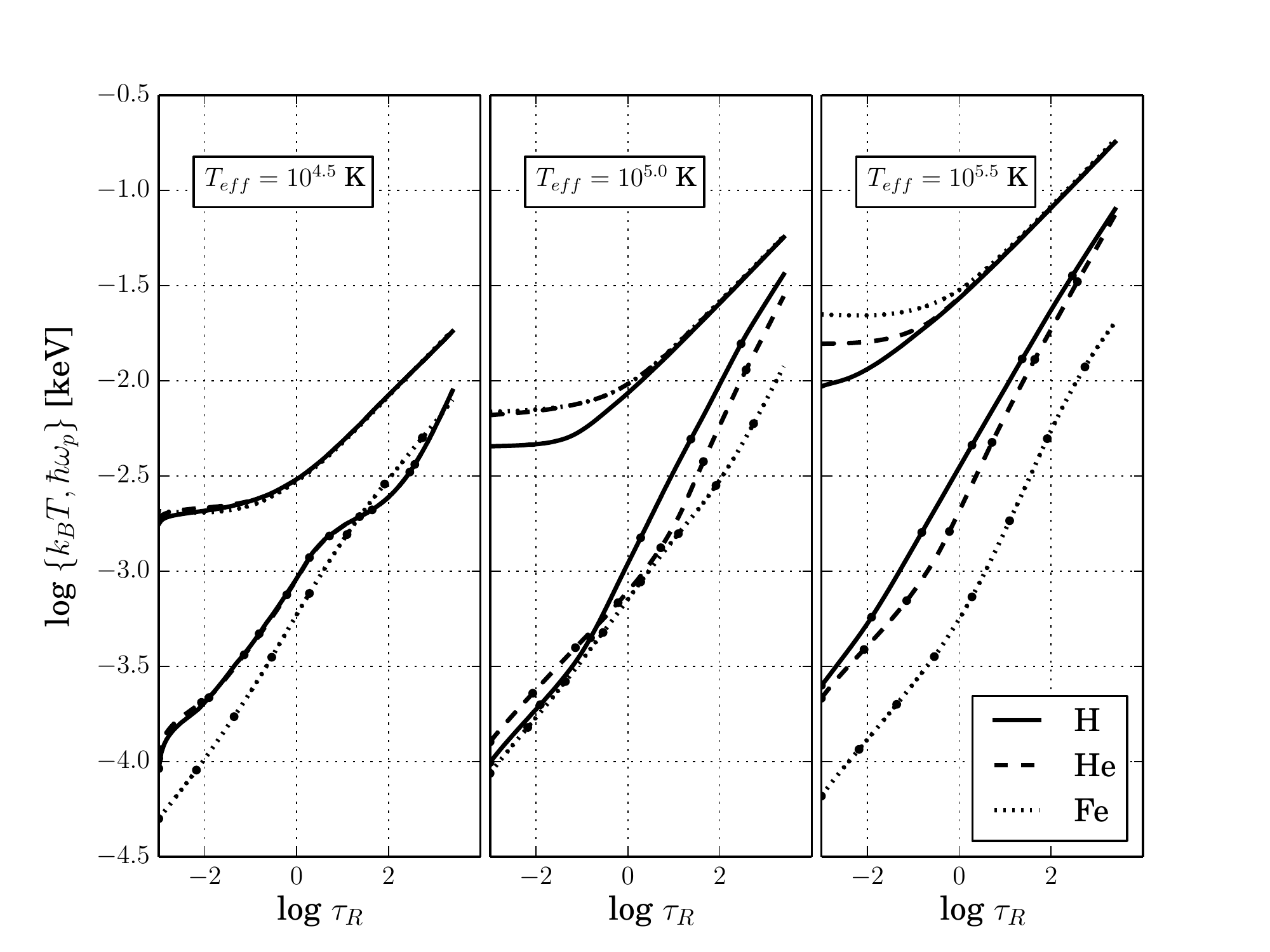}
\caption{Thermal energy, $k_{\mathrm{B}}T$, and energy associated to the plasma frequency, $\hbar \omega_p$, through the atmosphere of a NS with effective surface temperature $\teff=\ee{4.5}$, $\ee{5.0}$ and $\ee{5.5}\K$. All curves correspond to an NS with mass $\mns=1.5\msun$ and radius $\rinfty=15\km$. The upper solid, dashed, and dotted lines are the temperature profiles for H, He, and Fe atmospheres, respectively. Correspondingly, the lower lines with big dots are the energy associated to the plasma frequency for each atmosphere composition.}
\label{fig:Pefftau}
\end{figure}

Figure~\ref{fig:test} shows that the emergent spectra for H, He, and Fe composition calculated with both codes do not show substantial differences. In particular, for $\tinfeff=\ee{5.0}\K$, the fractional difference in the UV range, $-2.0\lesssim \log\left(E/\mathrm{keV}\right)\lesssim-1.3$, is always $<2\%$.  The largest differences likely originate in the width of the energy bins (wider in the work of \citealt{pons02}), which do not fully resolve the absorption lines. The agreement is much better in regions away from these lines. We also compare our Fe spectra with those of \cite{rajagopal96} 
finding no significant differences.

\begin{figure}
\centering
\includegraphics[width=\hsize, viewport=15 0 520 395,clip]{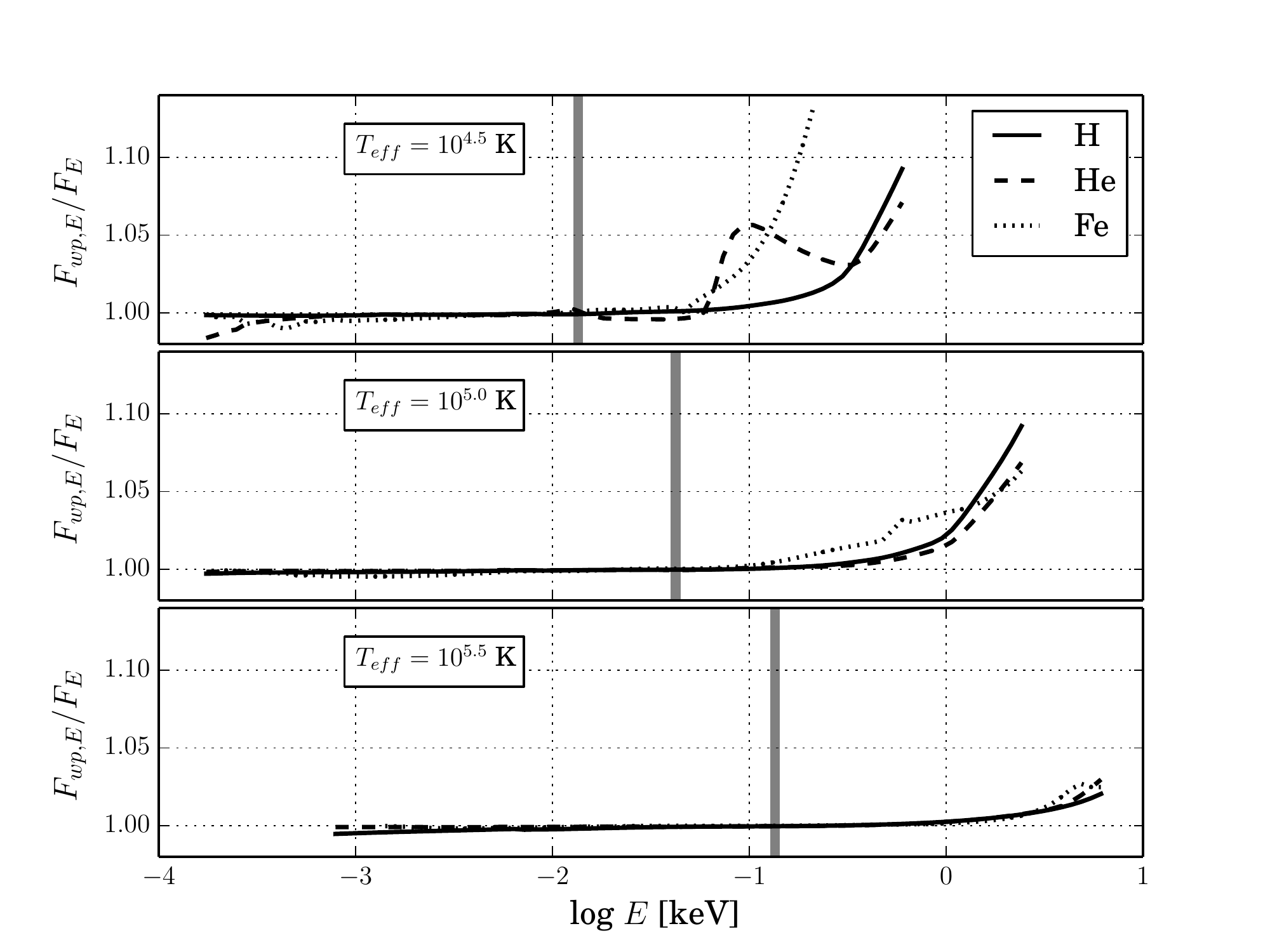} 
\caption{Ratio between fluxes for spectra with and without plasma frequency effects. The spectra are calculated considering $\mns=1.5\msun$, $\rinfty=15\km$, and effective surface temperatures $\teff=\ee{4.5}$,$\ee{5.0}$ and $\ee{5.5}\K$. The solid, dashed, and dotted lines correspond to H, He, and Fe atmospheres, respectively. As a reference,  the gray vertical lines indicate the energies where a black body with same $T_\mathrm{eff}$ has its maximal emission. The plasma  frequency has the largest impact on the high energy (Wien) tail of the  spectrum of the coldest Fe atmosphere model, but far from the flux peak.}
\label{fig:peff}
\end{figure}
\begin{figure}
\centering
\includegraphics[width=\hsize, viewport=30 0 550 395,clip]{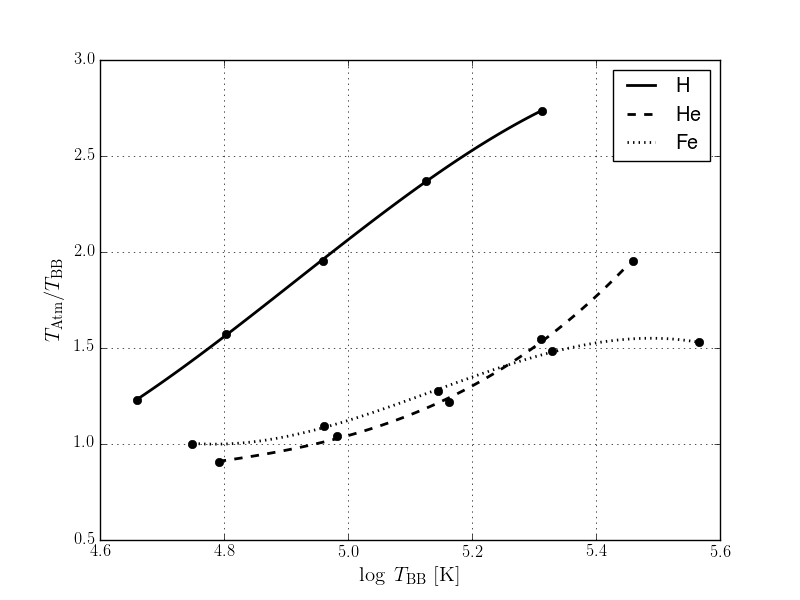} 
\caption{ Temperature transformation between BB and atmosphere spectral fits for ultraviolet observations, considering the energy band $E^\infty=6.2 - 9.4$~eV corresponding to the \hst\ F140LP filter. The solid, dashed, and dotted lines show polynomial functions fitted for the temperature transformation (in black points) in the cases of H, He, and Fe atmosphere models, setting $\mns=1.5\msun$ and $\rinfty=15\km$. For more details see the last paragraph of Section~\ref{sec:model}.}
\label{fig:const}
\end{figure}

We compute spectra for pure H, He, and Fe atmospheres with and without plasma effects.  The plasma frequency is obtained from the number of free electrons per nucleus, which is tabulated in the LANL opacity tables for a given composition, temperature, and density. Figure \ref{fig:Pefftau} shows that the energy associated with the plasma frequency, $\hbar\omega_p$, is always substantially below the peak of the spectra $\sim k_{B}T$. Since most of the flux is produced at energies $E\sim (1-10)k_{B}T$, plasma effects block an insignificant part of the photon flux. This means that plasma frequency effects do not produce a significant change in the temperature profile and the structure of the atmosphere. 

In addition, Figure \ref{fig:peff} shows that the plasma frequency effects change the spectra just slightly below $E\sim\ee{-3}\keV$ and above $E\sim\ee{-1}\keV$ for all effective surface temperatures considered.  The flux below $E\sim\ee{-3}\keV$ is reduced because low energy photons are blocked at relatively low Rosseland optical depths. Instead, the flux increases above $E\sim\ee{-1}\keV$ because, at high Rosseland optical depths, the plasma frequency blocks photons with higher energies, which, in order to conserve the radiative flux, requires a slight increase of the temperature in the inner parts of the atmosphere. In fact, since the opacities decrease at high energies, and the energy-dependent optical depth $\tau_{E} = 1$ is located at deeper zones, the emergent spectrum for $E\gtrsim\ee{-1}\keV$ is sensitive to the plasma effects and to the change of the temperature profile in the inner parts of the atmosphere. However, even with this change in the atmosphere spectra, the overall effect in the relevant  energy range $E\sim\ee{-2}-1\keV$ of the  thermal emission from J0437 (considering $\teff\sim\ee{5}\K$), is  $\lesssim2\%$ for H/He composition and  $\lesssim4\%$ for Fe composition. 

For the analysis presented in this section, we have considered only three representative effective surface temperatures, $\teff=\ee{4.5}$, $\ee{5.0}$, and $\ee{5.5}\K$. Figure~\ref{fig:peff} shows the tendency of plasma effects to become relatively less important, for all compositions, as the effective temperature of the atmosphere increases.  Therefore, for $\teff>\ee{5.5}\K$, plasma effects should be negligible. On the other hand, for $\teff<\ee{4.5}\K$, the amount of free electrons throughout the atmosphere decreases for all compositions, i.e., the plasma becomes less ionized. However, at low temperatures (as shown in Figure~\ref{fig:peff}), plasma frequency effects may still produce a substantial change in the spectra of Fe atmospheres  in the soft X--ray energy range,  although relatively far from the flux peak (as a reference, see also Figure~\ref{fig:test}). Thus, we conclude that plasma effects are not important in the thermal emission of MSPs (and any non-magnetic NSs) with light element atmosphere, but it may become important in very cold and  heavy element atmospheres,  producing   an enhancement of the soft X--ray emission\footnote{In the case of a liquid/metallic phase, at the surface of a very cold NS, the plasma frequency can be very high and suppress the flux in a substantial portion of the energy range of the thermal radiation. However, the radiative transfer equations are no longer valid for this regime, and the study of this case is outside the scope of this paper.}, but limited just to the high energy (Wien) tail of the spectrum.

Figure~\ref{fig:const} shows the relation between the temperature inferred from a BB fit and that from an atmosphere model, considering the flux in the energy range $E^\infty=6.2-9.4\eV$ (as observed at infinity). In particular, for a fixed ``redshifted radius'' or ``apparent radius'' $\rinfty=\left(1+z\right)\rns$ and temperature $T\sim10^5$~K, a BB fit  roughly reflects the effective temperature for  He and Fe atmospheres. Instead, a BB fit underestimates by a factor $\approx 2.1$ the  temperature with respect to a H  atmospheres.  In the range plotted, the temperature transformation between a BB and a H atmosphere can be fairly well described by the function $T_\mathrm{H}/T_\mathrm{BB} \equiv f_\mathrm{H} = -1.994 x^3 + 29.42 x^2 - 142.1 x + 226.4$, where $x= \log( T_\mathrm{BB}[\K])$. Similarly, the temperature transformations from BB to He and Fe atmospheres can be described by the functions $f_\mathrm{He}= 1.674 x^3 - 23.5 x^2 + 110.4 x - 172.6$ and $f_\mathrm{Fe} = -3.157 x^3 + 48.65 x^2 - 248.7 x + 423.2$, respectively.

\section{Applications to PSR~J0437--4715}
\label{sec:applications}

\subsection{Fitting procedure, MCMC and tests}
\label{subsec:MCMC}

We fit our spectral model of non-magnetic NS atmospheres to the UV and soft X--ray emission from J0437. For the UV band, we use spectroscopic and photometric data from \hst\ observations by \cite{kargaltsev04} and \cite{durant12}. For the soft X--ray band, we use \rosat\ archive data \citep{becker93} reanalyzed by \citet{guillot16a}. We compute a \chisq\ statistic between spectral models and UV/X--ray data considering four fitting parameters: interstellar extinction, $E(B-V)$, neutral hydrogen column density, \nh, effective temperature, $\tinfeff$, and radius, \rns, of the NS.

X--ray spectral fits account for the effects of interstellar neutral H via the absorption model of \citet{wilms00}. We fold the (absorbed) spectral model using the response matrix and effective area of the \rosat-PSPC camera, taken from the HEASARC webpage\footnote{\url{https://heasarc.gsfc.nasa.gov/docs/rosat/pspc_matrices.html}}. The folded X--ray spectra are binned in such a way that they match the energy binning of the ROSAT data presented in \citet{guillot16a}. CCD pile-up is not considered in our analysis as the X--ray data for J0437 show relatively low photon count-rates. To compute the \chisq\ statistic in the X--ray band, we considered only the $\sim 0.1 - 0.4\keV$ range, which is consistent with the cool thermal emission from the whole NS surface. In most of our analysis, we neglect the small contribution from the hot polar caps \citep{bogdanov13,guillot16a}, whose effect we evaluate approximately in \S~\ref{subsec:caps}.

We account for the dust effects in the UV fits using Milky Way extinction curves from \citet{clayton03}, which are computed using the polynomial function from \citet{fitzpatrick90}, setting $R_V = 3.1$. Following \citet{durant12}, we compute the \chisq\ statistic in the UV band considering the $7 - 11~\eV$ range of \hst\ data, which is consistent with a Rayleigh-Jeans tail of the surface emission. As discussed in \citet{durant12}, the spectrum of J0437 shows, just below $\sim 7\eV$, a peaked optical/UV excess whose origin is still unknown (where the instrument spectral response is also rapidly decreasing).

\begin{figure*}
    \centering
    \begin{subfigure}{0.49\textwidth}
        \includegraphics[width=\textwidth]{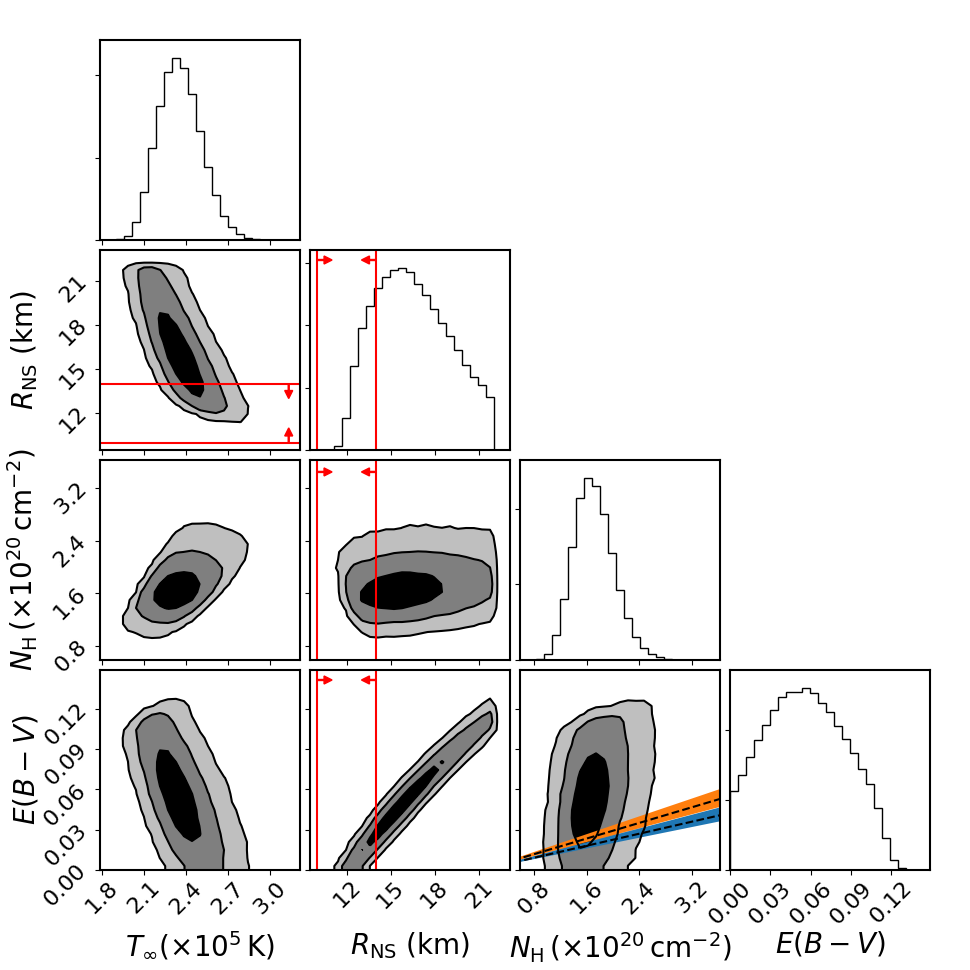}
        \caption{Hydrogen}
        \label{fig:Hfit}
    \end{subfigure}
    \begin{subfigure}{0.49\textwidth}
        \includegraphics[width=\textwidth]{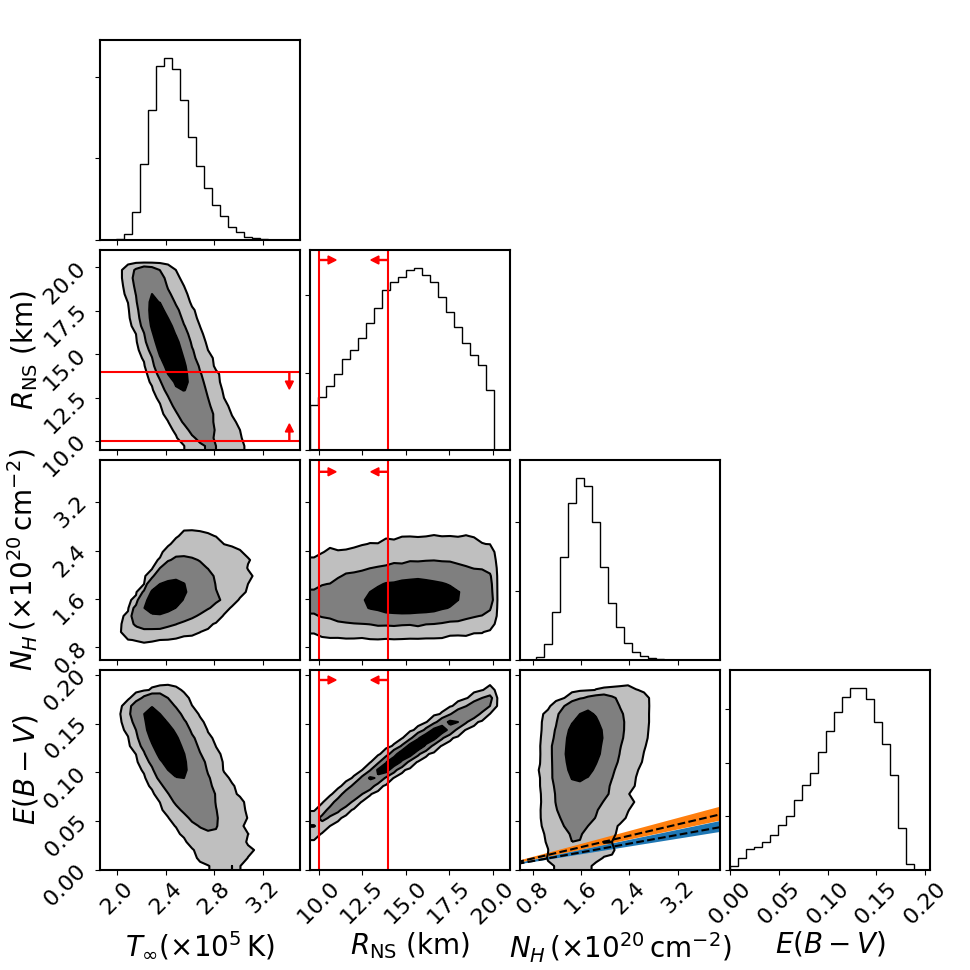}
        \caption{Helium}
        \label{fig:Hefit}
    \end{subfigure}
    \begin{subfigure}{0.49\textwidth}
        \includegraphics[width=\textwidth]{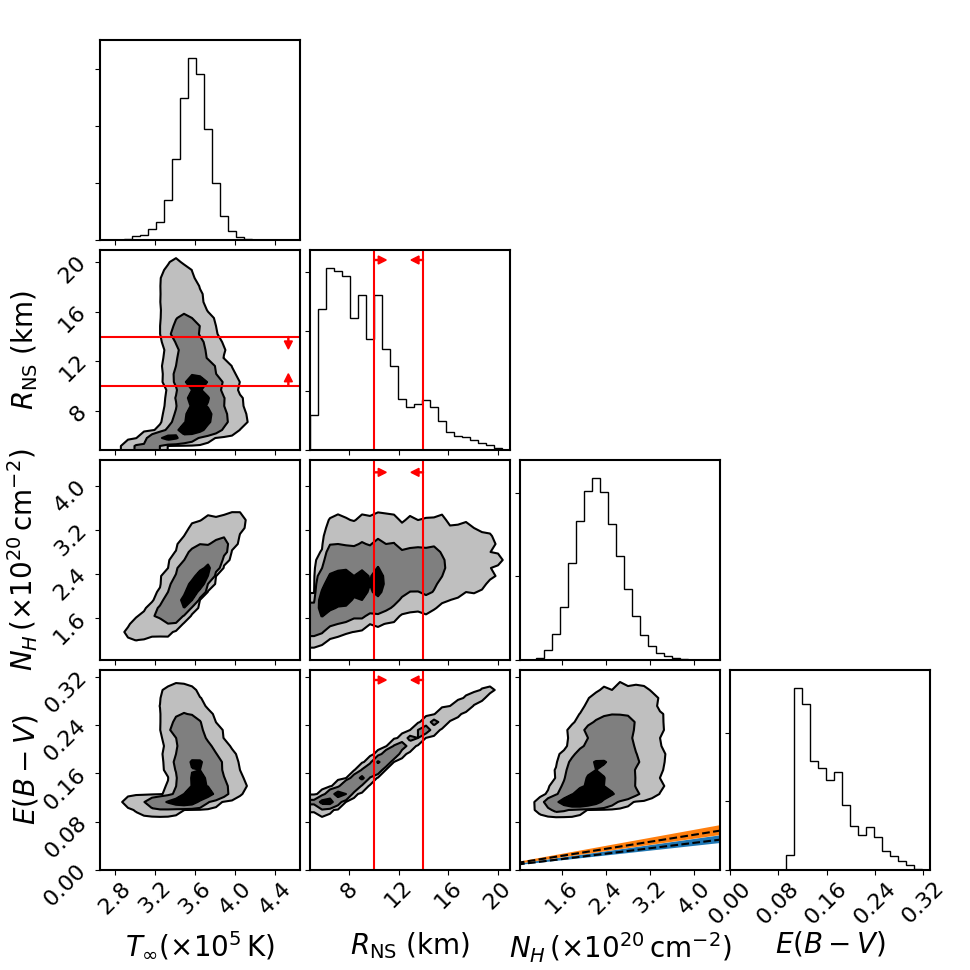}
        \caption{Iron}
        \label{fig:Fefit}
    \end{subfigure}
    \begin{subfigure}{0.49\textwidth}
        \includegraphics[width=\textwidth]{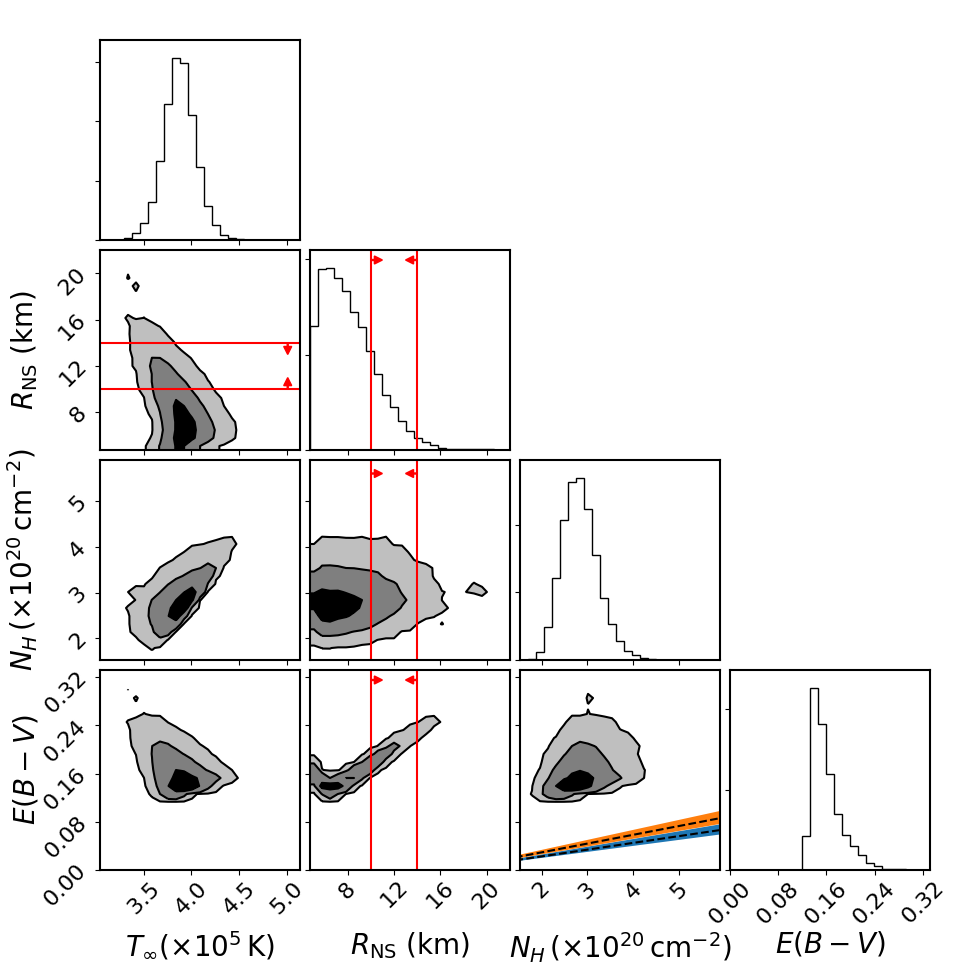}
        \caption{Blackbody}
        \label{fig:BBfit}
    \end{subfigure}
    \caption{
    MCMC marginalized (one-dimensional and two-dimensional) posterior distributions for the fitting parameters used in the spectral analysis of J0437. The models are computed for non-magnetized, partially ionized NS atmospheres considering H, He and Fe composition, as well as BB emission (panels a--d). (For the H atmosphere model, the Lyman alpha absorption line was eliminated, as explained in Section~\ref{sec:magnetic}.) The black, dark gray and gray regions show the $68\%$, $95\%$ and $99.7\%$ confidence levels. The orange and blue bands show the $E(B-V)$--$\nh$ relations ($3 \sigma$ enclosure around the central value as dashed lines) derived by \citet{guver09} and \citet{foight16}, respectively, for the Milky Way's interstellar medium. The red, vertical lines with the arrows show the $99\%$ confidence range of the constraints on the NS radius obtained from the gravitational wave event GW~170817  \citep{abbott18}. While all compositions and the BB model produce sensible measurements of the NS properties, the H atmosphere model produces the best agreement with the empirical interstellar $E(B-V)$--$\nh$ relations and $E(B-V)$ values from Galactic dust maps. 
    }
    \label{fig:allmcmc}
\end{figure*}

\begin{figure*}
\centering
\includegraphics[width=8.60cm,viewport=0 0 420 310,clip]{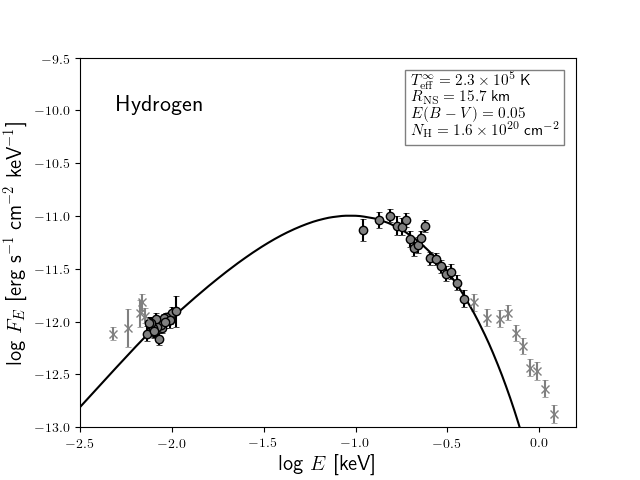}
\includegraphics[width=7.55cm,viewport=51 0 420 310,clip]{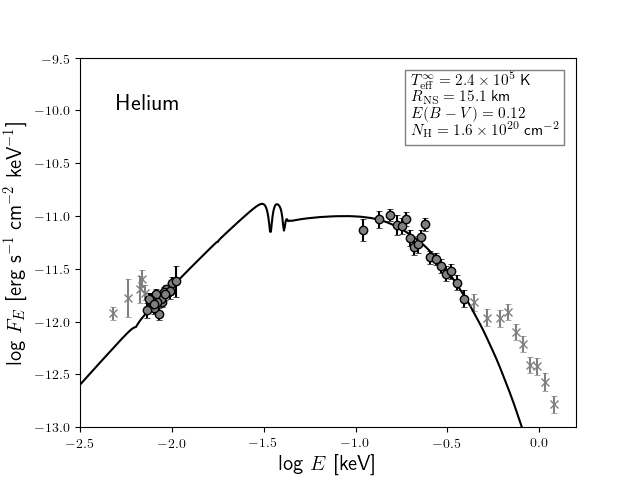}
\includegraphics[width=8.60cm,viewport=0 0 420 310,clip]{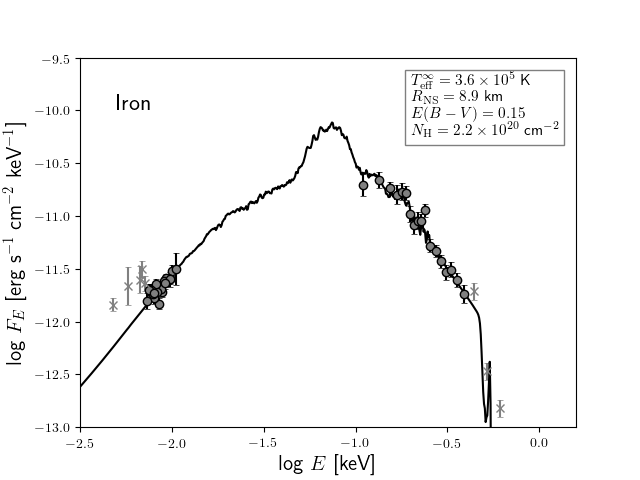}
\includegraphics[width=7.55cm,viewport=51 0 420 310,clip]{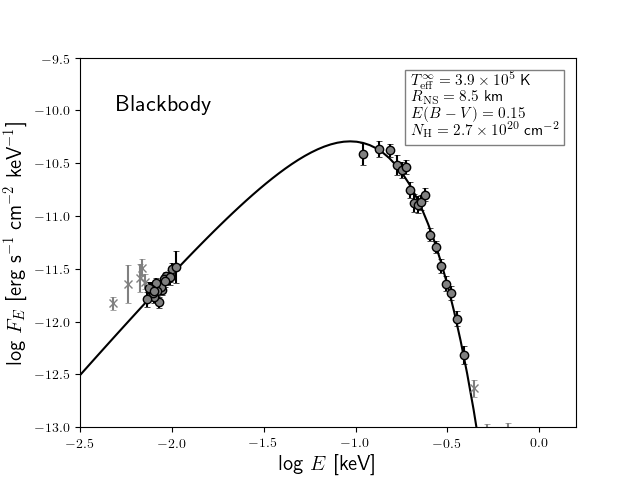}
\caption{Best fitting spectral models to the UV (\hst) and soft X--ray (\rosat) emission from J0437. In each panel, the solid lines show the spectral models computed for non-magnetized, partially ionized NS atmospheres considering H, He, and Fe composition, as well as BB emission. The gray filled circles and cross symbols with error bars show the data used and not used in the spectral fits, respectively (for more details see Section \ref{subsec:MCMC}).  The UV data correspond to de-reddened fluxes, and the soft X--ray data correspond to unfolded, unabsorbed fluxes. Since the unfolding process depends on the assumed models, specially at the higher energies, and the best-fitting values for the extinction and hydrogen column density are also different, the plotted data points differ from one panel to another. The best-fit values for $\tinfeff$, $\rns$, $E(B-V)$ and $\nh$ listed in each panel correspond to the peak of the posterior distribution shown in Figure~\ref{fig:allmcmc}, whereas those listed in Table~\ref{tab:fit} are the posterior medians.
}
\label{fig:MSPatm}
\end{figure*}

To obtain the confidence levels for the fitted parameters, we run a set of MCMC simulations using the package {\tt EMCEE} \citep{foremanmackey13}, considering four cases of emission models: H, He, and Fe atmosphere spectra and BB emission. The atmosphere spectra are obtained with log-scale polynomial interpolation from a $10\times10$ grid of models computed (from radiative transfer calculations, see Section~\ref{sec:theo}) in a suitable range of temperatures and radii.  We checked that the relative difference between interpolated spectra and actual atmosphere spectra is negligible, and we also found that our results remain largely unchanged by using for example a $5\times5$ temperature-radius grid. The models account for gravitational redshift considering a NS mass $\mns = 1.44\msun$, and setting the source distance to $d=156.79\pc$ \citep{reardon16}.

\begin{table}
    \begin{centering}
	\caption{Spectral fit parameters with different models for J0437.
	The values of  $\rns$, $\tinfeff$, $\nh$ and $E(B-V)$  are obtained from the MCMC posterior distributions. The values quoted are the medians (i.e., 50\% quantile), and the lower/upper uncertainties are obtained from the 16\% and 84\% quantiles, so that they provide the 68\% credible intervals.
	} 
	\resizebox{\columnwidth}{!}{
	\begin{tabular}{lcccll}\toprule 
		Model & \rns & $\tinfeff$ & \nh & $E(B-V)$ &$\chi^2/$dof  \\
        \noalign{\vskip 0.5mm}    
         & (km) &   ($\ee{5}$~K)   & $(\ee{20}\unit{\percmsq})$ &     \\
		\hline
        \noalign{\vskip 1mm}    
        H   & $16.3^{+3.0}_{-2.5}$ & $2.4^{+0.2}_{-0.2}$ & $1.7\pm0.3$ & $0.06^{+0.03}_{-0.03}$ & 43.7/34 \\
        \noalign{\vskip 2mm}    
        He  & $15.1^{+2.7}_{-3.2}$ & $2.5^{+0.2}_{-0.2}$ & $1.7\pm0.3$ & $0.12^{+0.03}_{-0.04}$ & 45.3/34\\
        \noalign{\vskip 2mm}    
        Fe  & $8.9^{+3.6}_{-2.4}$ & $3.6^{+0.2}_{-0.2}$ & $2.3\pm0.4$ & $0.15^{+0.06}_{-0.03}$ & 43.7/34 \\
        \noalign{\vskip 2mm}    
        BB  & $7.8^{+2.7}_{-1.9}$ & $3.9^{+0.2}_{-0.2}$ & $2.8\pm0.4$ & $0.15^{+0.03}_{-0.01}$ & 45.2/34\\
		\hline
        \multicolumn{6}{c}{Gaussian prior on $E(B-V)$}\\  
		\hline
        H  & $13.1^{+0.9}_{-0.7}$ & $2.5\pm0.1$ & $1.6\pm0.3$ & $0.01\pm0.01$ & 45.2/34\\
        \hline
        \multicolumn{6}{c}{Including hot polar caps and Gaussian prior on $E(B-V)$}\\  
		\hline
        H  & $13.6^{+0.9}_{-0.8}$ & $2.3\pm0.1$ & $1.4\pm0.3$ & $0.01\pm0.01$ & 45.9/34\\
		\bottomrule
	\end{tabular}
	}
	\label{tab:fit}
	\end{centering}
\end{table}

We first considered uniform prior distributions in all the fitting parameters ($\rns$, $\tinfeff$, $E(B-V)$, and $\nh$), and then we refined our results with a Gaussian prior on $E(B-V)$, with boundaries for the MCMC equal or larger than the limits shown in the Figure~\ref{fig:Hfit}--\ref{fig:BBfit}. For each MCMC run, we used 100 walkers (chains) over 10,000 iterations. The first 25\% iterations of each run were excluded when generating the posterior distributions. We also checked convergence of the MCMC by visual inspection of the traces of parameters, and of the likelihood (\chisq), to ensure the proper mixing and sampling of the parameter space. The final minimum \chisq\ were statistically acceptable with values between 43.7 and 45.9 depending on the model, for 34 degrees of freedom.

\begin{figure*}
\centering
\includegraphics[width=0.49\textwidth]{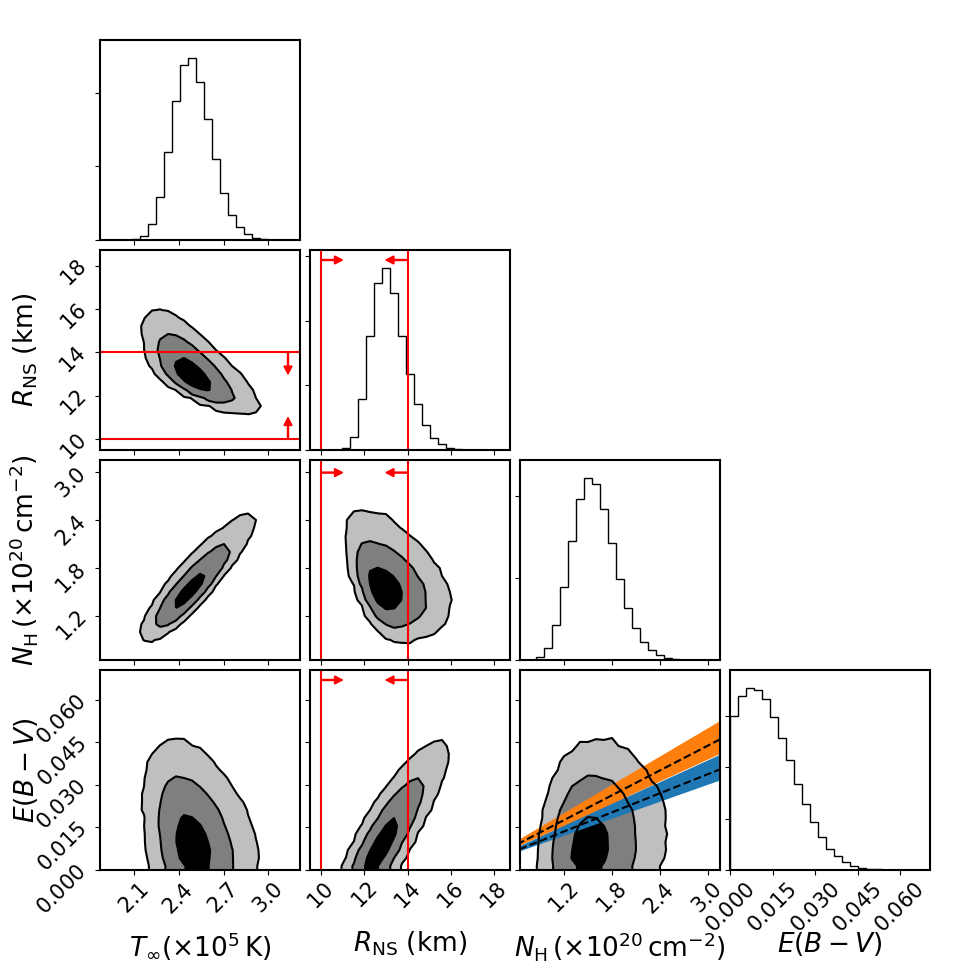}
\includegraphics[width=0.50\textwidth,viewport=0 0 420 403,clip]{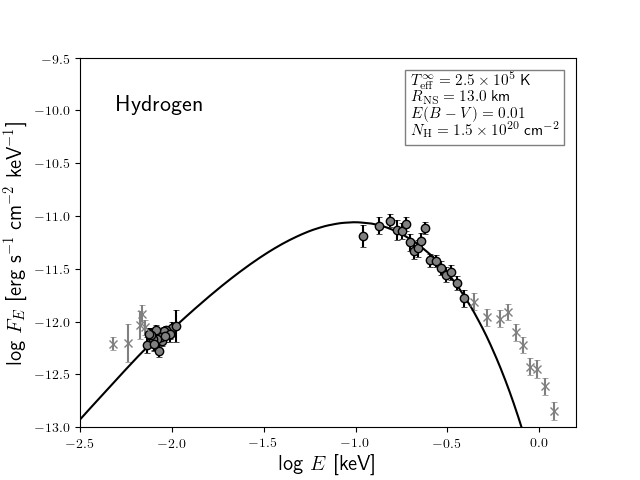}
\caption{Same as Figure~\ref{fig:allmcmc} and \ref{fig:MSPatm} for 
H atmosphere model, but considering a Gaussian prior on $E(B-V)$.}
\label{fig:Hprior}
\end{figure*}


\subsection{Results of spectral fits with uniform priors}
\label{sec:fits}

The results for the H, He, Fe, and BB spectral fits are summarized in Table~\ref{tab:fit}. The spectral fits for all emission models have equally good $\chi^2$ statistics. However, the results of our MCMC analyses, considering flat priors on all fitting parameters (Figure~\ref{fig:allmcmc}), suggest that the H atmosphere model is favoured, as its posterior distributions for $E(B-V)$ and $\nh$ show the best agreement with: 
\begin{itemize}
    \item[a)] the measurements $E(B-V)< 0.012$ obtained with a 2D map\footnote{\url{https://irsa.ipac.caltech.edu/applications/DUST/}} of infrared dust emission \citep[][see also \citealt{schlegel98}]{schlafly11} and $E(B-V)= 0.002\pm0.014$ (for distances $d =155-160$~pc to J0437) from a 3D map\footnote{\url{https://stilism.obspm.fr}} constructed from starlight absorption by dust \citep[see also \citealt{lallement14,capitanio17}]{lallement18},
    \item[b)] previous estimates of the interstellar dust extinction towards J0437 in the range $0.0<E(B-V)<0.07$ \citep{kargaltsev04,durant12}, and 
    \item[c)] the correlation between \nh\ and $A_V$ discussed by \citet[][see also \citealt{predehl95,guver13}]{foight16}: $\nh = (2.81\pm0.12)\times\ee{21} A_V$, where $A_V = E(B-V)\times R_V$ and $R_V$ is taken as 3.1.
\end{itemize}
Similar arguments suggest that the He atmosphere model is somewhat disfavoured, and the Fe atmosphere and BB models are ruled out.

\begin{figure*}
\centering
\includegraphics[width=0.49\textwidth]{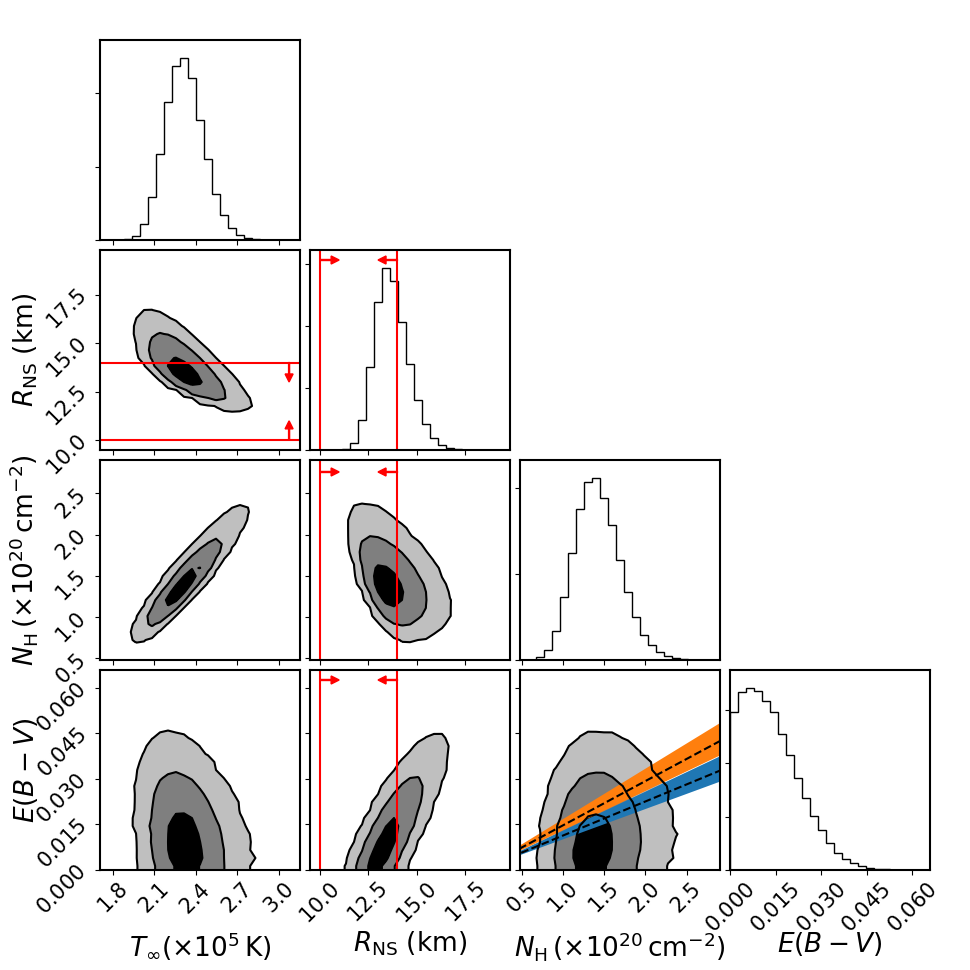}
\includegraphics[width=0.50\textwidth,viewport=10 0 530 400,clip]{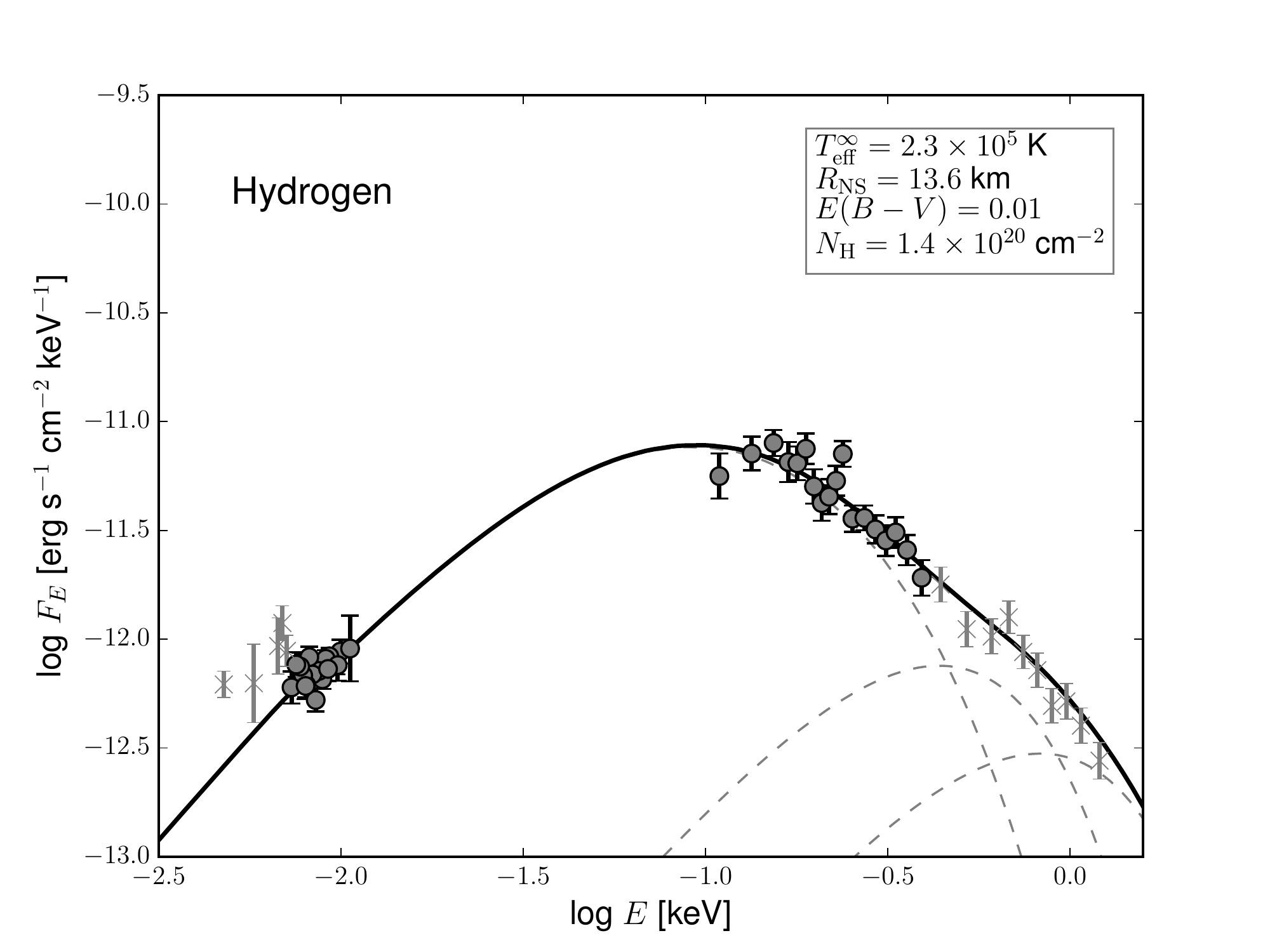}
\caption{Same as Figure~\ref{fig:Hprior}; for the H atmosphere model, considering a Gaussian prior on $E(B-V)$, but accounting for the emission from the hot polar caps. On the right panel, the black solid line corresponds to a H atmosphere model plus two hot blackbody models (for details see Section \ref{subsec:caps}). The gray, dashed lines show the contribution of each component to the model spectrum. The UV and X--ray data (filled circles and crosses) are defined as in Figure \ref{fig:MSPatm}.}
\label{fig:H_2BB_prior}
\end{figure*}

The posterior distributions for the H atmosphere model\footnote{Note that the constraints on the temperature and radius derived in our spectral fits are more restrictive than those reported by \citet{durant12}, which were obtained with combined UV (\hst) and soft \xray\ data (\xmmlong; spectral flux at $E=600\eV$ taken as an upper limit on the surface thermal emission), but considering a BB model and a range of NS radii between $\rns = 7\km$ and $\rns = 24\km$.} produce  a radius $\rns = 16.3^{+3.0}_{-2.5}\km$ and a bulk surface temperature $\tinfeff=\left(2.4\pm0.2\right)\tee{5}\K$, substantially cooler than the polar caps ($\gtrsim \ee{6}\K$). In particular, the radius measurement is consistent with the lower bound obtained from the analysis of the \xray\ light curve (hot polar cap emission) of J0437 by \citet[][]{bogdanov13}, or the analysis of the broad X--ray spectral shape \citep{guillot16a}. Furthermore, the lower end of the posterior distribution for $\rns$ is also compatible with the 99\% confidence limits on the NS radius obtained from the gravitational wave signal detected from the NS--NS merger GW~170817 (Figure \ref{fig:Hfit}, red lines), determined assuming a parameterized equation of state consistent with 1.97\msun\, \citep{abbott18}.

The posterior distributions obtained with all atmosphere models show a strong correlation between $\rns$ and $E(B-V)$ and an anti-correlation between $\tinfeff$ and $E(B-V)$. Therefore, an independent measurement of the dust extinction would strongly reduce the error intervals for the radius and temperature. If the extinction is negligible, as suggested by the Galactic dust maps mentioned above, the H atmosphere model can produce a NS radius as small as $\rns \sim 12$~km. A discussion of the MCMC analysis including a prior on $E(B-V)$ is given in Section \ref{subsec:prior}.

Figure \ref{fig:MSPatm} shows the best fitting spectra obtained with the MCMC analysis for all emission models.  The UV data are de-reddened according to the best-fit $E(B-V)$ and the soft X--ray data are unfolded\footnote{The soft X--ray data are unfolded as $X_{\textrm{unfold}}^{\textrm{data}} = X_{\textrm{fold}}^{\textrm{data}} \cdot S/S_{\textrm{folded}}$, where $X_{\textrm{fold}}^{\textrm{data}}$ correspond to the folded data, $S$ is the model spectrum, and $S_{\textrm{folded}}$ is the spectral model folded according to the telescope response.} and transformed to un-absorbed flux using the best fitted $\nh$. Since the posterior distributions for $\teff$, $\rns$, $E(B-V)$, and $\nh$ are non-Gaussian, the best-fit parameters shown in Figure \ref{fig:MSPatm} differ slightly from the posterior medians listed in Table~\ref{tab:fit}.

\subsection{NS radius estimation with a Gaussian prior on $E(B-V)$}
\label{subsec:prior}

J0437 is located in a region particularly devoid of dust\footnote{Rosine Lallement, private communication.}. As discussed in the previous subsection, 2D and 3D maps of dust extinction toward this source give $E(B-V) < 0.012$ and $E(B-V)= 0.002\pm0.014$, respectively. Furthermore, these values are compatible within $2\sigma$ with those derived with the MCMC analysis for the H atmosphere model (using flat priors in all the fitting parameters) and the empirical $\nh$--$E(B-V)$ relation \citep{foight16}.

We repeat the MCMC analysis for the H atmosphere model including a Gaussian prior on $E(B-V)$, with mean $\mu_\mathrm{dust} = 0.002$ and standard deviation $\sigma_\mathrm{dust} = 0.014$, according to the latest 3D map of Galactic dust \citep{lallement18}, while ensuring $E(B-V)>0$. A summary with the resulting medians for the fitting parameters is reported in Table~\ref{tab:fit}. As expected, the results are compatible with those reported in Section \ref{sec:fits}. Remarkably, the uncertainties on the radius measurement are substantiallyreduced, yielding $\rns = 13.1^{+0.9}_{-0.7}\km$ (see  also the posterior distributions in Figure \ref{fig:Hprior}).

\subsection{Correction for hot polar caps and final radius estimate}
\label{subsec:caps}

Up to this point, we have neglected the effects on our fits from the hot polar caps, which are clearly identified in the X-ray data above $\sim 0.5\unit{keV}$ \citep{becker93,pavlov97,bogdanov13,guillot16a}. In our low-energy spectral analysis, there could be two such effects:
\begin{itemize}
    \item[a)] The low-energy tail of the hot components could directly contribute to the high end of the spectral range considered in our models, and
    \item[b)] the folded soft X--ray spectrum can be contaminated with high-energy photons due to the spectral response of the detector.
\end{itemize}

We tested the effect of the hot polar caps by adding two hot BB components to our H atmosphere fit (see Figure \ref{fig:H_2BB_prior}). We used the parameters for the polar caps obtained by \citet[][]{guillot16a} for their H atmosphere + 2BB fit, which includes \nustar, \xmmlong, and  \rosat\ data, covering the X--ray spectrum of J0437 up to 20 \keV.  Their best fitting temperatures for the polar caps are $T_{\mathrm{cap},1}^\infty=1.8\times10^6$~K and $T_{\mathrm{cap},2}^\infty=3.4\times10^6$~K, with associated radii $R_{\mathrm{cap},1}^\infty=0.15$~km and $R_{\mathrm{cap},2}^\infty=0.03$~km, respectively. By including these fixed components in our MCMC analysis, we obtain a final NS radius estimation   $\rns=13.6^{+0.9}_{-0.8}$~km (a summary with all  posterior medians  is reported in Table \ref{tab:fit}). In comparison to the results reported in Section \ref{subsec:prior}, the addition of the hot components makes the inferred bulk temperature decrease and the NS radius increase by amounts smaller than the estimated 1 $\sigma$ error bars.

A full analysis, including atmosphere model fits of the emission from both the hot polar caps and the cooler surface of the rest of the NS, is outside the scope of this paper and should be addressed using, for example, the hard X-ray data from \nustar\ and  the high-quality soft X--ray data from the \nicer{} mission. Furthermore, dealing with the hot polar cap emission  requires to properly account for scattering opacity in the source function of our atmosphere models, which at the moment are suitable to model the cold thermal component of J0437 (see Section \ref{sec:theo}).   Other uncertainties, such as the errors in the source distance or mass, are negligible in our analysis, as the errors are dominated by the much larger uncertainties, for example, in the interstellar extinction $E(B-V)$ (set as a fitting parameter in the MCMC analysis, either with a uniform or Gaussian prior).

\begin{figure}
    \centering
    \includegraphics[width=0.50\textwidth,viewport=20 5 550 400,clip]{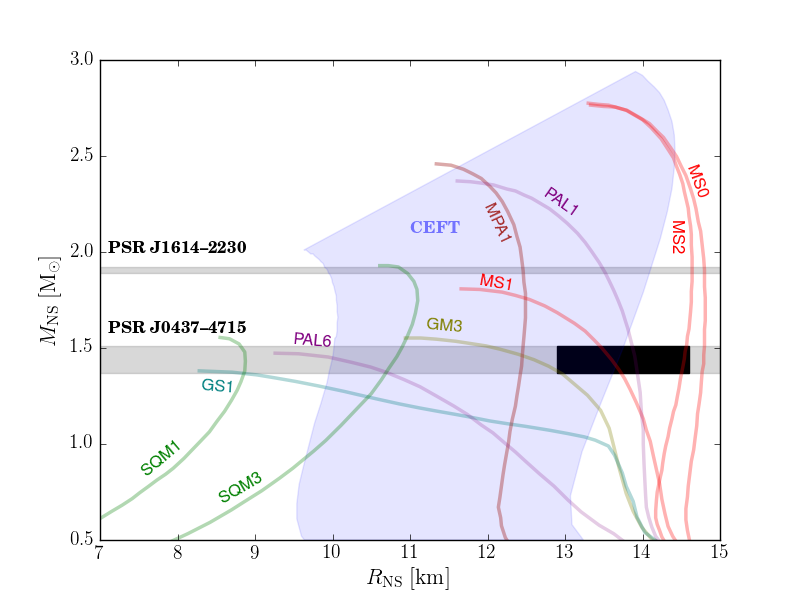}
    \caption{Mass-radius relation for different cold, superdense matter EOSs. The  curves with different colors show a few  EOSs, labeled as in  \citet[][]{lattimer01}. The blue filled region labeled ``CEFT'' shows a range of EOSs based on chiral effective field theory \citep{hebeler13}. The grey,  horizontal bands show the mass measurements for PSR J1614--2230 \citep{demorest10,arzoumanian18} and  J0437 \citep{reardon16}. The  black region shows  the radius measurement for J0437, at $1\sigma$, obtained in this work.
     }
    \label{fig:EOS}
\end{figure}

\section{Discussion and conclusions}
\label{sec:ccl}

We modelled the cool thermal component of the spectrum of
MSP J0437, observed in the UV (\hst) and soft X--ray (\rosat) bands, considering  non-magnetized, partially ionized H, He, and Fe atmospheres. For surface temperatures $\sim10^5$~K, as previously determined for this source \citep{kargaltsev04,durant12}, we found that plasma effects are negligible in the UV band (less than $1\%$ flux suppression), but may become important for cooler sources with heavy-element atmospheres, particularly for the emission in the soft X--ray band (in the Wien tail of the spectrum).

 Using a MCMC analysis, we found that spectral fits to the UV/X--ray data of J0437 favour a H atmospheric composition, disfavour a He composition, and rule out Fe atmospheric composition as well as BB emission.  This is consistent with the fact that BB emission cannot reproduce the observed pulsed amplitude of J0437 \citep{bogdanov13}. For the H atmosphere composition, we found that:
 \begin{itemize}
     \item[a)] By considering uniform priors in all fitting parameters, we obtain a NS radius $\rns=16.3^{+3.0}_{-2.5}\km$, a bulk surface temperature $\tinfeff=\left(2.4\pm0.2\right)\tee{5}\K$, a dust extinction value $E(B-V)=0.06\pm0.03$, and a neutral H column density $\nh=\left(1.7\pm0.3\right)\tee{20}\percmsq$.
     \item[b)] By including a Gaussian prior on the dust extinction, based on current 3D maps of galactic dust, we refine our measurements: $\rns=13.1^{+0.9}_{-0.7}\km$, $\tinfeff=\left(2.5\pm0.2\right)\tee{5}\K$, and $\nh=\left(1.6\pm0.3\right)\tee{20}\percmsq$
     \item[c)] By accounting for the effect of the hot polar caps, we obtain our final results: $\rns = 13.6^{+0.9}_{-0.8}\km$,  $\tinfeff=(2.3\pm0.1)\tee{5}\K$, and  $\nh=\left(1.4\pm0.3\right)\tee{20}\percmsq$
 \end{itemize}

Our radius determination for J0437, combined with its well-measured mass, allows us to establish the tightest constraint on the equation of state for ultra-dense matter to date \cite[for a review see][]{lattimer16} from a MSP. As shown in Figure~\ref{fig:EOS}, the constraint on \mns\,\citep{reardon16} and \rns\,(this work) for J0437 combined with one of the largest measured masses for a pulsar \citep[PSR J1614--2230,][]{demorest10,arzoumanian18} favours a stiff EOS and disfavours a strange matter EOS.  Precise 3D maps of Galactic dust, presently under development, based on \emph{GAIA} data \citep[see e.g.,][]{lallement19}, and high quality X--ray observation from the \nicer{} mission, will further improve the radius estimation for J0437 and the constraints on the EOS.

 Compared with other results, our  measurement of \rns\  for J0437 is:
\begin{itemize}
\item[a)]  consistent with the lower limits on the radius  previously published for this source. Specifically, \cite{bogdanov13} derived $\rns>10.9\km$ from the X--ray light curve (due to the hot polar caps), assuming a H atmosphere, while \cite{guillot16a} obtained $\rns>10\km$ from the soft X--ray spectrum (using a BB spectral component for the cool surface),  
\item[b)]  consistent with the constraints derived for two other MSPs: PSRs J2124 and J0030, with the associated lower limits $\rns> 7.8\km$ and $\rns> 10.4\km$, respectively \citep[assuming $\mns=1.4~\Msun$,][]{bogdanov08,bogdanov09},
\item[c)]  consistent with the NS radius  derived from the NS-NS merger gravitational wave signal GW~170817 \citep{abbott18}, 
\item[d)]  consistent with the NS radius measurement from recent statistical analyses combining  quiescent low-mass X--ray binaries \citep[e.g.,][]{steiner18,baillot19}, which find radii in the 11--14\km\ range, and
\item[e)] slightly larger, but still marginally consistent with the NS radius  obtained  through the analysis of the cooling  tails of   X--ray  bursts   from the  low-mass  X--ray binary 4U 1702--429,  $\rns=12.4\pm0.4\km$  \citep{nattila17}.
\end{itemize}

Our analysis also allowed us to test the surface composition of the MSP J0437. In particular, a H atmosphere is in agreement with the expectations for these sources, as such composition might result from a) past accretion from a binary companion, b) accretion from the interstellar medium or c) spallation of heavier elements \citep{bildsten92}. If other heavier elements coexist in the surface layers of NSs, they would stratify within $\sim100$~s \citep{romani87,bildsten92}, leaving the lightest element on top. Furthermore, a very small amount of H, $\sim\ee{-20}\msun$ \citep{bogdanov16}, is enough to produce an optical depth $\sim 1$ in the X--rays (i.e. to form the atmosphere). On the other hand, the favoured H atmospheric composition of J0437 disfavours diffuse nuclear burning, or at least its effectiveness in modifying the atmosphere composition on $\sim\ee{4}\yr$ timescales \citep{chang03,chang04,chang10} and produce external atmospheric layers with heavier elements. Systematic observations of the UV and soft X--ray emission from other MSPs may help us to establish further constraints on this issue.

The bulk surface temperature of J0437 derived in the present work with the H atmosphere model is more restrictive than that obtained with a BB model by \citet[][$\tinfeff=(1.5-3.5)\times10^5\K$]{durant12}, which does not take into account the minimal value of $R_\infty=3\sqrt{3}\mathrm{G}M/\mathrm{c}^2 = 11.0\pm0.5$~km imposed by General Relativity (considering the currently measured mass the pulsar $\mns=1.44\pm0.07\msun$, \citealt{reardon16}). 
Our temperature measurement is also relevant to understand the heating mechanisms that might be operating in NSs. \cite{gonzalez10} performed a comparative analysis of different heating mechanisms, finding that rotochemical heating \citep{reisenegger95,reisenegger97,fernandez05,petrovich10,petrovich11,gonzalezjimenez15} and vortex creep \citep{alpar84,shibazaki89,larson99} might explain the temperatures measured in old NSs. Rotation-induced crustal heating, which was proposed later \citep{gusakov15}, could also be important. A full analysis of this issue will be presented in another work (Rodr\'iguez et al., in prep.).

We note that our results rely on the assumption that non-magnetized atmosphere models appropriately describe the thermal emission from the entire surface of relatively cold NSs, such as J0437. The spin-down-derived magnetic field of this object, $B=2.4\times10^8\G$, can affect the transport of radiation in the atmospheric plasma for electromagnetic waves with energies lower than the electron cyclotron energy  $E_\mathrm{c}/(1+z)\sim 2 \eV$, which is still below the UV band considered in our fits. Potentially, small-scale multipolar components present on the NS surface (stronger than the dipolar field) could affect the radiative transfer. If that is the case, and assuming that the transport of radiation becomes polarized (propagating in the so called X-mode and O-mode), this would produce an excess in the optical/UV spectrum compared with the non-magnetized atmosphere model \citep[see e.g.,][]{ho01,zane01,lloyd03,suleimanov12}. Such spectral fits will produce a smaller NS radius, and so our results with non-magnetized atmosphere models may be considered as upper limits. A further caveat to consider is the possibility that no atmosphere is present on the cold surfaces of NSs (but probably at much lower surface temperatures than that of J0437), where the emission would arise directly from a liquid surface. However, no models exist so far to describe such emission from sources like MSPs.


\section*{Acknowledgements}

We thank Jos\'e Pons for providing us with NS spectra used to test of the present models and Rosine Lallement, George Pavlov, Andrew Cumming, Crist\'obal Espinoza, M\'arcio Catelan, Thomas Tauris, Silvia Zane, Kinwah Wu, Mat Page, Ignacio Ferreras, Paul Kuin, Nicol\'as Viaux, and the ANSWERS group in Santiago for useful discussions that helped improve this paper. This research was supported by FONDECYT Regular Grant 1171421, ECOS-CONICYT project C16U01 and CONICYT Basal project AFB-170002. D.G.-C. acknowledges the financial support of \emph{Becas Chile} fellowship No. 72150555. S.G. acknowledges the support of the French Centre National d'\'{E}tudes Spatiales (CNES).

\bibliographystyle{mnras}
\bibliography{biblio}

\bsp	
\label{lastpage}
\end{document}